\documentclass{aa}
\usepackage{graphicx,amsfonts,amssymb,txfonts,natbib,latexsym}
\newcommand{\plotwd}{8.5cm}

\begin{document}

\titlerunning{SDSS LRG cosmological parameters}
\title{Power spectrum of the SDSS luminous red galaxies: constraints on cosmological parameters}
\author{G. H\"utsi\inst{1,2}}

\institute{Max-Planck-Institut f\"ur Astrophysik, Karl-Schwarzschild-Str. 1, 86740 Garching bei M\"unchen, Germany
\and Tartu Observatory, T\~oravere 61602, Estonia}

\offprints{G.H\"utsi , \\ \email{gert@mpa-garching.mpg.de}}
\date{Received / Accepted}

\abstract
{In this paper we determine the constraints on cosmological parameters using the CMB data from the {\sc Wmap} experiment together with the recent power spectrum measurement of the SDSS Luminous Red Galaxies (LRGs). Specifically, we focus on spatially flat, low matter density models with adiabatic Gaussian initial conditions. The spatial flatness is achieved with an additional quintessence component whose equation of state parameter $w_\mathrm{eff}$ is taken to be independent of redshift. Throughout most of the paper we do not allow any massive neutrino contribution and also the influence of the gravitational waves on the CMB is taken to be negligible. The analysis is carried out separately for two cases: (i) using the acoustic scale measurements as presented in \citet{astro-ph/0512201}, (ii) using the full SDSS LRG power spectrum and its covariance matrix. We are able to obtain a very tight constraint on the Hubble constant: $H_0 = 70.8 ^{+2.1}_{-2.0}\,\mathrm{km/s/Mpc}$, which helps in breaking several degeneracies between the parameters and allows us to determine the low redshift expansion law with much higher accuracy than available from the {\sc Wmap} + HST data alone. The positive deceleration parameter $q_0$ is found to be ruled out at $5.5\sigma$ confidence level. Finally, we extend our analysis by investigating the effects of relaxing the assumption of spatial flatness and also allow for a contribution from massive neutrinos.
}
\maketitle

\section{Introduction}
Since the flight of the {\sc Cobe} \footnote{http://lambda.gsfc.nasa.gov/product/cobe/} satellite in the beginning of 90's the field of observational cosmology has witnessed an extremely rapid development. The data from various Cosmic Microwave Background (CMB) experiments ({\sc Wmap} \footnote{http://map.gsfc.nasa.gov/} \citep{2003ApJS..148....1B}, {\sc Cobe} \citep{1992ApJ...396L...1S}, {\sc Archeops} \footnote{http://www.archeops.org/} \citep{2003A&A...399L..19B}, {\sc Boomerang} \footnote{http://cmb.phys.cwru.edu/boomerang/} \citep{2002ApJ...571..604N}, {\sc Maxima} \footnote{http://cfpa.berkeley.edu/group/cmb/} \citep{2000ApJ...545L...5H}, {\sc Cbi} \footnote{http://www.astro.caltech.edu/$\sim$tjp/CBI/} \citep{2003ApJ...591..556P}, {\sc Vsa} \footnote{http://www.mrao.cam.ac.uk/telescopes/vsa/} \citep{2003MNRAS.341.1076S}, {\sc Dasi} \footnote{http://astro.uchicago.edu/dasi/} \citep{2002ApJ...568...38H} etc.), supernova surveys ({\sc Scp} \footnote{http://supernova.lbl.gov/} \citep{1999ApJ...517..565P},  High-Z SN Search \footnote{http://cfa-www.harvard.edu/oir/Research/supernova/HighZ.html} \citep{1998AJ....116.1009R}) and large galaxy redshift surveys (SDSS \footnote{http://www.sdss.org/} \citep{2000AJ....120.1579Y}, 2dFGRS \footnote{http://www.mso.anu.edu.au/2dFGRS/} \citep{2001MNRAS.328.1039C}) has lead us to the cosmological model that is able to accommodate almost all the available high quality data-- the so-called ``concordance'' model \citep{1999Sci...284.1481B,2003ApJS..148..175S}. Useful cosmological information has also been obtained from the Ly-$\alpha$ forest, weak lensing, galaxy cluster, and large-scale peculiar velocity studies. It is remarkable that this diversity of observational data can be fully explained by a cosmological model that in its simplest form has only $5-6$ free parameters \citep{2004MNRAS.351L..49L,2004PhRvD..69j3501T}. As the future data sets will be orders of magnitude larger, leading to the extremely small statistical errors, any further progress is possible only in case we fully understand various systematic uncertainties that could potentially bias our conclusions about the underlying cosmology. As such, one should try to use observables that are least sensitive to the theoretical uncertainties, contaminating foregrounds etc. Currently the ``cleanest'' constraints on cosmological models are provided by the measurements of the angular power spectrum of the CMB. Since the underlying linear physics is well understood (see e.g. \citealt{astro-ph/9508126,2003moco.book.....D}) we have a good knowledge of how the angular position and amplitude ratios of the acoustic peaks depend on various cosmological parameters. However, the CMB data alone is able to provide accurate measurements of only a few combinations of the cosmological parameters. In order to break the degeneracies between the parameters one has to complement the CMB data with additional information from other independent sources e.g. the data from the type Ia supernovae, large-scale structure, or the Hubble parameter measurements. In fact, the well understood physical processes responsible for the prominent peak structure in the CMB angular power spectrum are also predicted to leave imprints on the large-scale matter distribution. Recently the analysis of the spatial two-point correlation function of the Sloan Digital Sky Survey (SDSS) Luminous Red Galaxy (LRG) sample \citep{2005ApJ...633..560E}, and power spectra of the 2dF \citep{2005MNRAS.362..505C} and SDSS LRG \citep{astro-ph/0512201} redshift samples, have lead to the detection of these acoustic features, providing the clearest support for the gravitational instability picture, where the large-scale structure of the Universe is believed to arise through the gravitational amplification of the density fluctuations laid down in the very early Universe. 

In the current paper we work out the constraints on cosmological parameters using the SDSS LRG power spectrum as determined by \citet{astro-ph/0512201}. In order to break the degeneracies between the parameters we complement our analysis with the data from other cosmological sources: the CMB data from the {\sc Wmap}, and the measurement of the Hubble parameter by the HST Key Project \footnote{http://www.ipac.caltech.edu/H0kp/}. We focus our attention on simple models with Gaussian adiabatic initial conditions. In the initial phase of the analysis we further assume spatial flatness, and also negligible massive neutrino and gravitational wave contributions. This leads us to the models with 6 free parameters: total matter and baryonic matter density parameters: $\Omega_m$ and $\Omega_b$, the Hubble parameter $h$, the optical depth to the last-scattering surface $\tau$, the amplitude $A_s$ and spectral index $n_s$ of the scalar perturbation spectrum. \footnote{In fact, one might even consider a simpler case with only 5 free parameters by fixing $n_s=1$ \citep{2004MNRAS.351L..49L}.} This minimal set is extended with the constant dark energy effective equation of state parameter $w_{\mathrm{eff}}$. We carry out our analysis in two parts. In the first part we use only the measurement of the acoustic scale from the SDSS LRG power spectrum as given in \citet{astro-ph/0512201}. The analysis in the second part uses the full power spectrum measurement along with the covariance matrix as provided by \citet{astro-ph/0512201}. Here we add two extra parameters: bias parameter $b$ and parameter $Q$ that describes the deformation of the linear power spectrum to the nonlinear redshift-space spectrum. These extra parameters are treated as nuissance parameters and are marginalized over. Thus the largest parameter space we should cope with is 9-dimensional. \footnote{Since marginalization over the bias parameter can be done analytically \citep{2002MNRAS.335.1193B} the actual number of parameters can be reduced to 8.} Since the parameter space is relatively high dimensional it is natural to use Markov Chain Monte Carlo (MCMC) techniques. For this purpose we use publicly available cosmological MCMC engine {\sc Cosmomc}\footnote{http://cosmologist.info/cosmomc/} \citep{2002PhRvD..66j3511L}. All the CMB spectra and matter transfer functions are calculated using the fast Boltzmann code {\sc Camb}\footnote{http://camb.info/} \citep{2000ApJ...538..473L}. 

The paper is organized as follows. In Sect. 2 we describe the observational data used for the parameter estimation. Sect. 3 discusses and tests the accuracy of the transformations needed to convert the linear input spectrum to the observed redshift-space galaxy power spectrum.  In Sect. 4 we present the main results of the cosmological parameter study and we conclude in Sect. 5.                  

\section{Data}
\begin{figure}
\centering
\includegraphics[width=\plotwd]{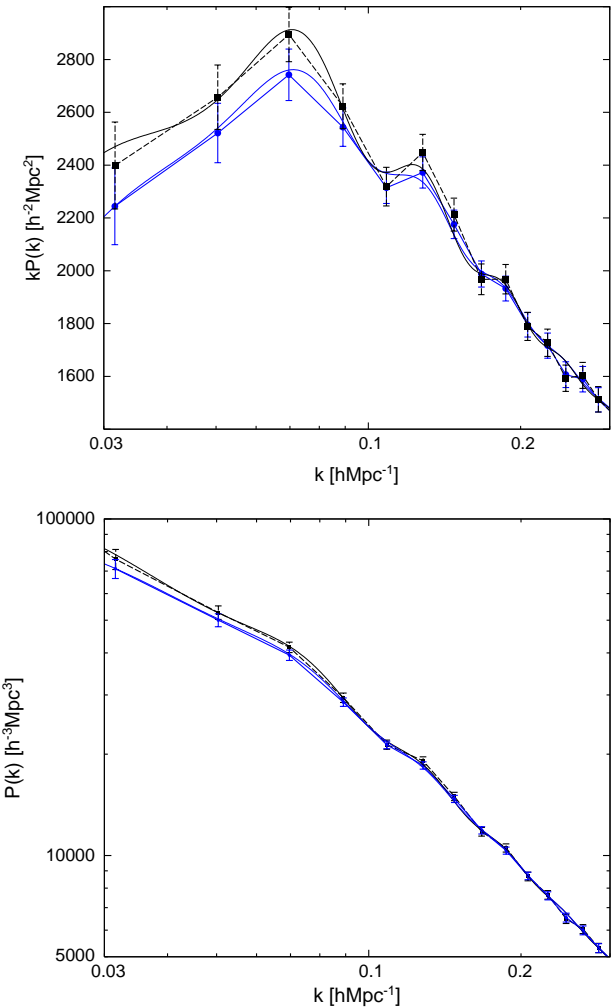}
\caption{ Upper panel: Power spectra in somewhat unconventional form. Here the spectra have been multiplied by an extra factor of $k$ to increase the visibility of details. Filled circles with solid errorbars represent the SDSS LRG power spectrum as determined by \citet{astro-ph/0512201}. The upper data points provide the deconvolved version of the spectrum. The thin solid lines show the best-fitting model spectra. Lower panel: The same spectra as above now plotted in the usual form.}
\label{fig1}
\end{figure}

The SDSS LRG power spectrum as determined by \citet{astro-ph/0512201} is shown with filled circles and heavy solid errorbars in Fig. \ref{fig1}. There the upper data points correspond to the deconvolved version of the spectrum. \footnote{The deconvolution was performed using an iterative algorithm due to \citet{1974AJ.....79..745L} with a specific implementation as given in \citet{1996ApJ...471..617L}.} The thin solid lines represent the best-fitting model spectra, with the lower curve corresponding to the convolved case. As the survey window is relatively narrow the deconvolution can be done rather ``cleanly''. This deconvolved spectrum might be useful for the extra-fast parameter estimation employing analytic approximations for the matter transfer functions \citep{1998ApJ...496..605E,1999A&A...347..799N} and fast CMB angular power spectrum generators such as CMBfit\footnote{http://www.hep.upenn.edu/~sandvik/CMBfit.html}\citep{2004PhRvD..69f3005S}, DASh\footnote{http://bubba.ucdavis.edu/DASh/}\citep{2002ApJ...578..665K} and CMBwarp\footnote{http://www.physics.upenn.edu/~raulj/CMBwarp/}\citep{2004PhRvD..70b3005J}. However, in this paper, as we use an accurate Boltzmann solver {\sc Camb} to calculate CMB power spectra and matter transfer functions, the relative time taken by an extra convolution step is completely negligible. Thus in the following we use only the convolved spectrum. \footnote{Often also called a pseudospectrum.} Accurate analytic fitting formulae for the survey window functions can be found in \citet{astro-ph/0512201}.\footnote{There the combination 'mode coupling kernels' is used in place of the more common 'window functions'.} The power spectrum covariance matrix in \citet{astro-ph/0512201} was measured from $1000$ mock catalogs generated with the second-order optimized Lagrangian perturbation calculation. The same paper also provides the measurement of the acoustic scale: $(105.4 \pm 2.3)\,h^{-1}\,\mathrm{Mpc}$. This corresponds to the case when only sinusoidal modulation, as expected in the case of adiabatic initial conditions, in the power spectrum is allowed. Relaxing this assumption by allowing an arbitrary phase shifts gave the result, $(103.0 \pm 7.6)\,h^{-1}\,\mathrm{Mpc}$, instead. In the following parameter estimation process we use both of these values. In \citet{astro-ph/0512201} the measurement of the acoustic scale was achieved by first removing the ``smooth'' component of the spectrum and then fitting the parametrized family of functions to the oscillatory part via the modified version of the Levenberg-Marquardt method.  The separation of the ``smooth'' and ``oscillatory'' components of the spectrum can be done rather accurately since the characteristic scales over which they change differ strongly. The Levenberg-Marquardt method which was used to determine the oscillation frequency approximates the likelihood surface near its maximum with a multidimensional Gaussian, and this way provides an approximate parameter covariance matrix. To avoid this ``Gaussianity assumption'' we have also performed a MCMC parameter estimation exercise, finding the best fitting acoustic scale along with its uncertainty in full agreement with the values quoted above. The question that might arise of course is how adequate is the parametric family that was used for fitting the oscillatory component? Even in the simplest case of the adiabatic initial fluctuations the damped sinusoidal modulation is only an approximation. We investigate the possible biases introduced by assuming a fixed parametric form for the oscillatory part of the spectrum in more detail in Sect. \ref{sect4}.

As mentioned in the Introduction, in order to break several degeneracies between the cosmological parameters, we complement the SDSS LRG power spectrum data with the data from the {\sc Wmap} CMB measurements. Specifically, we use the CMB temperature power spectrum as found in \citet{2003ApJS..148..135H} and the temperature-polarization cross-power as determined by \citet{2003ApJS..148..161K}. The description of the likelihood calculation using this data is given in \citet{2003ApJS..148..195V}. We use the Fortran90 version of this likelihood code as provided by the {\sc Cosmomc} package.

While investigating the constraints arising from the measurement of the acoustic scale we do not run each time the full new MCMC calculation. Instead we importance sample the chains built for the {\sc Wmap} data along with the constraint on the Hubble parameter as provided by the HST Key Project, $H_0=72 \pm 8\,\mathrm{km/s/Mpc}$ \citep{2001ApJ...553...47F}. Using the {\sc Wmap} data alone would result in too loose constraints on several parameters, and thus after importance sampling a large fraction of the chain elements would get negligible statistical weight, leaving us with too small effective number of samples.     

\section{Power spectrum / acoustic scale transformation}\label{sect3}
In this section we discuss the relation of the observed galaxy power spectrum to the underlying spectrum of the matter distribution. We stress the need to take into account the so-called cosmological distortion \footnote{In order to convert the observed redshifts to the comoving distances needed for the estimation of the power spectrum, one has to assume some background cosmological model -- the fiducial cosmology. If the true underlying cosmology differs from the fiducial model we are left with a distortion of the power spectrum, which is often called the cosmological distortion.}, which almost always is being completely neglected. \footnote{According to our knowledge the only counter-example being the work by \citet{2005ApJ...633..560E}.} This is fine for the very shallow surveys, but as we show later, for the samples like the SDSS LRGs, with an effective depth of $z_{\mathrm{eff}}\sim 0.35$, the cosmological distortion should certainly be taken into account. This is especially important if power spectrum, instead of being well approximated by a simple power law, contains some characteristic features. 

There are other difficulties one has to face while trying to make cosmological inferences using the observed galaxy samples. It is well known that galaxies need not faithfully follow the underlying matter distribution. This phenomenon is known as biasing \citep{1984ApJ...284L...9K}. Whereas on the largest scales one might expect linear and scale-independent biasing (e.g. \citealt{1993MNRAS.262.1065C,2000ApJ...528....1N}), on smaller scales this is definitely not the case. In general the biasing can be scale-dependent, nonlinear, and stochastic \citep{1999ApJ...520...24D}. The other complications involved are the redshift-space distortions and the effects due to nonlinear evolution of the density field. The redshift-space distortions, biasing, and nonlinearities can be approximately treated in the framework of the Halo Model approach as described in Appendix \ref{appa}. The implementation of the Halo Model as presented there introduces four new parameters: $M_{\mathrm{low}}$, the lower cutoff of the halo mass i.e. below that mass halos are assumed to be ``dark''; $\alpha$ and $M_0$, the parameters describing the mean of the halo occupation distribution i.e. the average number of galaxies per halo with mass $M$, which was assumed to have the form $\langle N|M \rangle = \left ( \frac{M}{M_0} \right )^\alpha$; $\gamma$, the parameter describing the amplitude of the virial motions inside the haloes with respect to the isothermal sphere model. This formulation of the Halo Model, along with the assumption of the best-fit {\sc Wmap} cosmology \citep{2003ApJS..148..175S}, is able to produce a very good fit to the observed SDSS LRG power spectrum as demonstrated in Fig. \ref{fig1}. Moreover, all the parameters: $M_{\mathrm{low}}$, $\alpha$, $M_0$, $\gamma$, are reasonably well determined. It turns out that to a good approximation these four extra parameters can be compressed down just to a single parameter $Q$, describing the deformation of the linearly evolved spectrum:
\begin{equation}\label{eq1}
P_{\mathrm{gal}}(k)=b^2(1+Qk^{\eta})P_{\mathrm{lin}}(k)\,.
\end{equation}
\begin{figure}
\centering
\includegraphics[width=\plotwd]{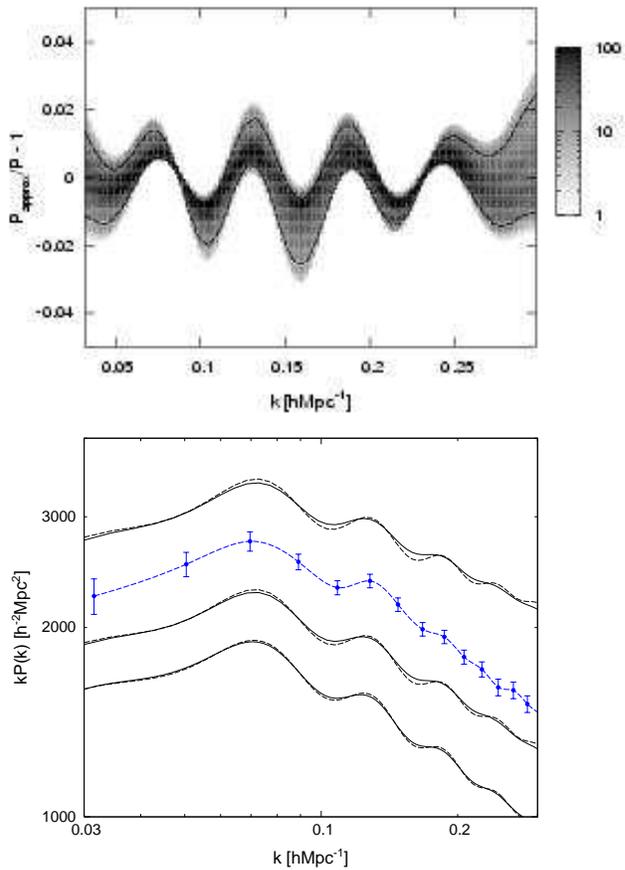}
\caption{Upper panel: A density plot showing the probability distribution functions for the relative accuracy of the approximation given in Eq. (\ref{eq1}) with $\eta=\frac{3}{2}$. The set of Halo Model parameters $M_{\mathrm{low}}$, $\alpha$, $M_0$, and $\gamma$, needed to calculate the ``exact'' spectra, were drawn from the multidimensional Gaussian distribution centered at the best-fit values and with a covariance matrix as found in \citet{astro-ph/0512201}. The heavy dashed lines mark the $5\%$ and $95\%$ quantiles of the relative accuracy distributions. Lower panel: Filled circles with solid errorbars provide the SDSS LRG power spectrum. The data points are connected with a smooth cubic spline fit. The other set of lines represents some examples of the pairs of spectra that correspond, starting from below, to the best matching case, to the $68\%$, and to the $90\%$ quantiles of the distribution of the $\chi^2$ values. The solid lines show the Halo Model spectra while the dashed ones are the approximations from Eq. (\ref{eq1}).}
\label{fig2}
\end{figure}
\begin{figure}
\centering
\includegraphics[width=\plotwd]{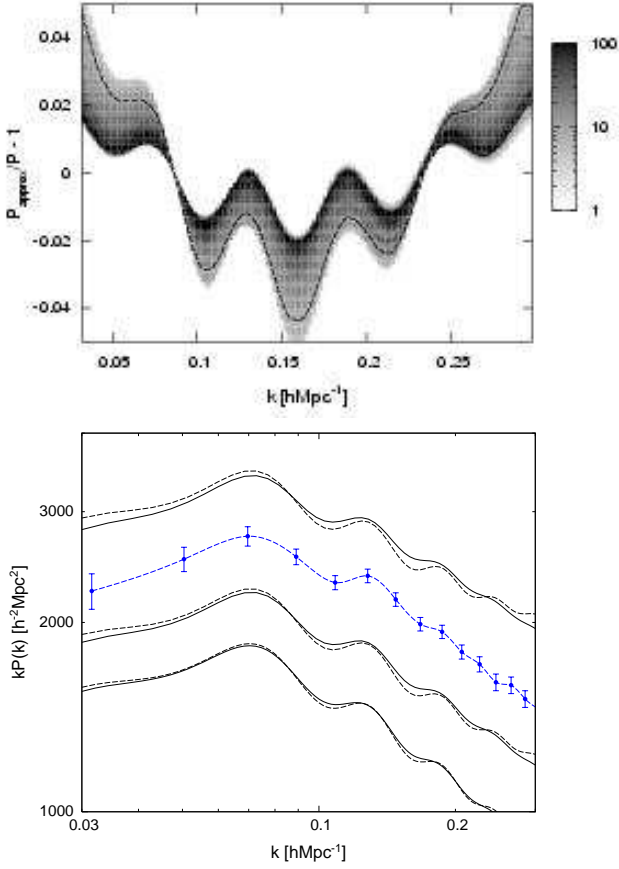}
\caption{The version of Fig. \ref{fig2} with $\eta=2$.}
\label{fig3}
\end{figure}
Here $b$ is the bias parameter and a good value for $\eta$ turns out to be $\frac{3}{2}$. A similar type of parametric description for the galaxy power spectrum was also used in \citet{2005MNRAS.362..505C}, with a slight difference for the treatment of the largest scales. In that paper the authors suggest to take $\eta=2$. However, we have found that $\eta=\frac{3}{2}$ provides a better approximation for these 4-parameter Halo Model spectra. \footnote{At least if the spectra have shapes close to the observed SDSS LRG spectrum.} This is demonstrated in the upper panels of Figs. \ref{fig2} and \ref{fig3}. There we have calculated a full range of Halo Model spectra (assuming the {\sc Wmap} ``concordance'' cosmology) for different values of $M_{\mathrm{low}}$, $\alpha$, $M_0$, and $\gamma$, drawn from the multidimensional Gaussian centered around the best-fitting values and with the parameter covariance matrix as found in \citet{astro-ph/0512201}. Each of the calculated models is fitted with a simple parametric form as given in Eq. (\ref{eq1}). The upper panels of Figs. \ref{fig2} and \ref{fig3} represent the density plots for the fractional accuracy of these simple fits i.e. for each wavenumber $k$ they show the probability distribution functions for the achieved relative accuracy. With the heavy dashed lines we have also marked the $5\%$ and $95\%$ quantiles of the accuracy distributions. It is evident that $\eta=\frac{3}{2}$ provides significantly better approximation to the spectral deformation than $\eta=2$. The largest errors are seen to be located at the positions of the acoustic features, with a simple approximation in Eq. (\ref{eq1}) giving larger oscillation amplitudes. The Halo Model gives lower oscillation amplitudes since relatively flat contribution from the 1-halo term added to the 2-halo part starts to decrease the contrast of the acoustic features, whereas the multiplicative transform in Eq. (\ref{eq1}) preserves the contrast level of these wiggles. In the lower panels of Figs. \ref{fig2} and \ref{fig3} we have provided some examples of the pairs of spectra that correspond (staring from below) to the best matching case, and also the ones representing the $68\%$ and $90\%$ quantiles of the distribution of the $\chi^2$ values. The solid lines here correspond to the Halo Model calculations. For comparison also the SDSS LRG power spectrum along with the cubic spline fitted to the data points are shown. For clarity slight vertical shifts have been applied to the model spectra. As can be seen, the approximation in Eq. (\ref{eq1}) is rather acceptable in the light of the accuracy of the SDSS LRG power spectrum measurement. This approximation is used in Sect. \ref{sect5} where we fit the model spectra to the SDSS LRG data. 

\begin{figure}
\centering
\includegraphics[width=\plotwd]{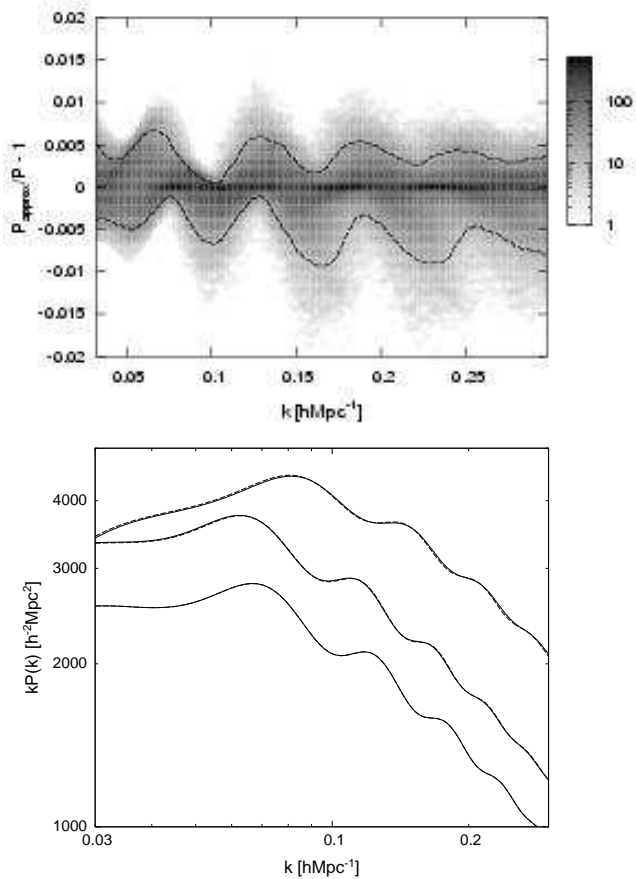}
\caption{An analog of Fig \ref{fig2}, here provided in the context of the accuracy test for the cosmological distortion approximation given in Eq. (\ref{eq12}). The set of cosmological models was drawn from the combined posterior corresponding to the {\sc Wmap} plus HST Key Project data.} 
\label{fig4}
\end{figure}

\begin{figure}
\centering
\includegraphics[width=\plotwd]{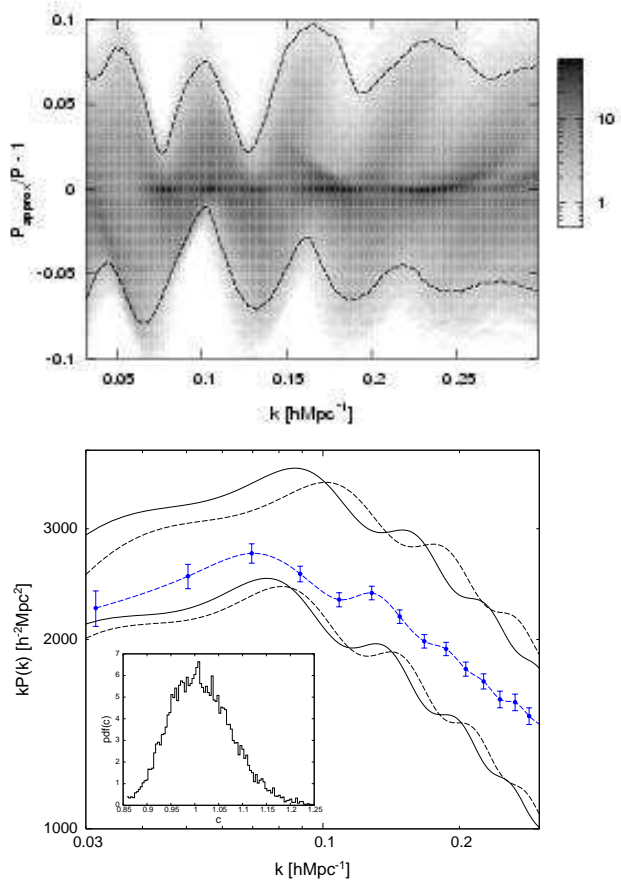}
\caption{As Fig. \ref{fig4}, here instead showing the error one makes if cosmological distortion is completely neglected. In the lower panel we have shown only the examples corresponding to the $68\%$ and $90\%$ quantiles. The inset shows the probability distribution function for the ``isotropized'' dilation scale, as given in Eq. (\ref{eq11}), compatible with the {\sc Wmap} plus HST Key project constraints.}
\label{fig5}
\end{figure}

The cosmological distortion, mentioned in the beginning of this section, arises due to the simple fact that conversion of the observed redshifts to comoving distances requires the specification of the cosmological model. If this cosmology differs from the true one, we are left with additional distortion of distances along and perpendicular to the line of sight. In general, the spatial power spectrum measurements, in contrast to the angular spectra, are model dependent i.e. along with the measurements of the 3D power spectrum one always has to specify the so-called fiducial model used to analyze the data. The fiducial model corresponding to the data shown in Fig. \ref{fig1} is the best-fit {\sc Wmap} ``concordance'' model \citep{2003ApJS..148..175S}. In principle, for each of the fitted cosmological model one should redo the full power spectrum analysis to accommodate different distance-redshift relation. However, there is an easier way around: one can find an approximate analytical transformation that describes how the model spectrum should look like under the distance-redshift relation given by the fiducial model i.e. instead of transforming the data points we transform the fitted model spectra. Since the distance intervals along and perpendicular to the line of sight transform differently, the initial isotropic theoretical spectrum $P$ transforms to the 2D spectrum:
\begin{equation}  
\widetilde{P}^{2D}(k_{\parallel},k_{\perp};z)=\frac{1}{c_{\parallel}(z)\cdot c_{\perp}^2(z)}P\left[\sqrt{\left(\frac{k_{\parallel}}{c_{\parallel}(z)}\right)^2+\left(\frac{k_{\perp}}{c_{\perp}(z)}\right)^2};z\right]\,,
\end{equation}  
where the distortion parameters along and perpendicular to the line of sight are defined as:
\begin{eqnarray}
c_{\parallel}(z)&=&\frac{H^{\mathrm{fid}}(z)}{H(z)}\,,\\
c_{\perp}(z)&=&\frac{d_{\perp}(z)}{d_{\perp}^{\mathrm{fid}}(z)}\,.
\end{eqnarray}
Here $H(z)$ is the Hubble parameter and $d_{\perp}(z)$ is the comoving angular diameter distance corresponding to the fitted theoretical model. Superscript $^{\mathrm{fid}}$ refers to the fiducial model. Here and in the following we use a tilde on top of $P$ to denote theoretical spectrum ``transformed to the reference frame of the fiducial cosmology''. As we use the spectra that have the dimensions of volume an extra division by $c_{\parallel}(z)\cdot c_{\perp}^2(z)$ occurs due to the transformation of the volume elements:
\begin{equation} 
{\mathrm d}V(z)=c_{\parallel}(z)\cdot c_{\perp}^2(z)\cdot{\mathrm d}V^{\mathrm{fid}}(z)\,.
\end{equation} 
By introducing the variables
\begin{equation} 
k=\sqrt{k_{\parallel}^2+k_{\perp}^2}\,,\quad \mu = \frac{k_{\parallel}}{k}\,,\quad \varkappa(z) = \frac{c_{\parallel}(z)}{c_{\perp}(z)}\,,
\end{equation} 
we can express $\widetilde{P}^{2D}$ as follows:
\begin{eqnarray}
& & \widetilde{P}^{2D}(k,\mu;z)= \nonumber \\
& & \quad \frac{1}{c_{\parallel}(z)\cdot c_{\perp}^2(z)}P\left[\frac{k}{c_{\perp}(z)}\sqrt{1+\left(\frac{1}{\varkappa^2(z)-1}\right)\mu^2};z\right]\,.
\end{eqnarray}
Now the corresponding isotropized spectrum can be given as:
\begin{eqnarray}
& & \widetilde{P}(k;z)= \nonumber \\
& & \quad \frac{1}{2\,c_{\parallel}(z)\cdot c_{\perp}^2(z)}\int\limits_{-1}^{1}P\left[\frac{k}{c_{\perp}(z)}\sqrt{1+\left(\frac{1}{\varkappa^2(z)-1}\right)\mu^2};z\right]{\mathrm d}\mu\,.
\end{eqnarray}
As the observations are done along the light-cone we have to perform relevant integrals along the redshift. The full treatment for the light-cone effect can be found in \citet{1999ApJ...517....1Y,1999ApJ...527..488Y}. As we are investigating a two-point function, an accurate light-cone calculation would introduce two integrals over the redshifts \citep{1997MNRAS.286..115M,1999ApJ...527..488Y}. However, it turns out that to a good approximation, excluding the very largest scales, the contributions from different redshifts decouple and the double integral reduces to a simple one-dimensional integral over redshift. The final result for the $\widetilde{P}(k;z)$, averaged over the light-cone can be given as:
\begin{equation}\label{eq8}
\widetilde{P}(k)=\frac{\int\limits_{z_{\mathrm{min}}}^{z_{\mathrm{max}}}\frac{{\mathrm d}V^{\mathrm{fid}}}{{\mathrm d}z}{\mathrm d}z\cdot \mathcal{W}^2(k;z)\,\bar{n}^2(z)\,\widetilde{P}(k;z)\,c_{\parallel}(z)\,c_{\perp}^2(z)}{\int\limits_{z_{\mathrm{min}}}^{z_{\mathrm{max}}}\frac{{\mathrm d}V^{\mathrm{fid}}}{{\mathrm d}z}{\mathrm d}z\cdot \mathcal{W}^2(k;z)\,\bar{n}^2(z)\,c_{\parallel}(z)\,c_{\perp}^2(z)}\,.
\end{equation} 
Here the result of \citet{1999ApJ...527..488Y} has been generalized to include other weight factors in addition to the simple number density weighting. The most common weight functions $\mathcal{W}(z)$ are the following:
\begin{equation}\label{eq9}
\mathcal{W}(k;z)\propto \left\{ 
\begin{array}{lll}
\frac{1}{\bar{n}(z)} & \rm{\quad for\ volume\ weighting}\\
\rm{const} & \rm{\quad for\ number\ weighting}\\
\frac{1}{1+\bar{n}(z)\widetilde{P}(k;z)} & \rm{\quad for\ the\ FKP\ weighting\,.}
\end{array} \right.
\end{equation}
Here FKP stands for the weighting scheme due to \citet{1994ApJ...426...23F}. The power spectrum measurement of the SDSS LRGs in \citep{astro-ph/0512201} used the FKP weighting function. In Eqs. (\ref{eq8}) and (\ref{eq9}) $\bar{n}(z)$ represents the mean number density of galaxies as a function of redshift. For the SDSS LRG sample analyzed in \citet{astro-ph/0512201} the limiting redshifts $z_{\mathrm{min}}=0.16$ and $z_{\mathrm{max}}=0.47$. If instead of the integral over $z$ in Eq. (\ref{eq8}) we just take the integrand at the effective redshift (e.g. the median redshift) of the survey, and replace the distortion parameters $c_{\parallel}$ and $c_{\perp}$ with a single ``isotropized'' dilation of scales (see e.g. \citealt{2005ApJ...633..560E}):
\begin{equation}\label{eq11}  
c_{\mathrm{isotr}}=\sqrt[3]{c_{\parallel}(z_{\mathrm{eff}})\,c_{\perp}^2(z_{\mathrm{eff}})}\,,
\end{equation}  
we can write instead of Eq. (\ref{eq8}) simply
\begin{equation}\label{eq12}  
\widetilde{P}(k)=\frac{1}{c_{\mathrm{isotr}}}P\left(\frac{k}{c_{\mathrm{isotr}}}\right)\,.
\end{equation}
Here the prefactor $1/c_{\mathrm{isotr}}$ can also be dropped, as it can be absorbed into the bias parameter that is assumed to be a completely free parameter throughout this paper.   
Although the true transformation for the power spectrum is different along and perpendicular to the line of sight, and also is dependent on redshift, it turns out that a single dilation approximation taken at the median redshift of the survey can provide a very good approximation, especially for relatively shallow surveys. For the median redshift of the SDSS LRG sample as analyzed in \citet{astro-ph/0512201}, $z\sim 0.35$, this approximation is very accurate as can be seen in Fig. \ref{fig4}. The upper panel of Fig. \ref{fig4} shows a similar density plot as in Figs. \ref{fig2} and \ref{fig3}. Here, in comparison to Figs. \ref{fig2} and \ref{fig3} where the background cosmology was fixed to the best-fit {\sc Wmap} model and the Halo Model parameters were varied, we use the simple linear spectra while changing the cosmology. The set of cosmological models is drawn from the combined posterior corresponding to the {\sc Wmap} plus HST Key Project data. As can be seen from this figure, for relatively shallow surveys the single ``isotropized'' dilation approximation is very precise: for $\sim 90\%$ of the models the approximation in Eq. (\ref{eq12}) is more accurate than $0.5\%$. This is even more clear when looking at the lower panel of Fig. \ref{fig4} where we have plotted the pairs of spectra corresponding to the best matching case, and also some examples representing $68\%$ and $90\%$ quantiles of the distribution of the $\chi^2$ values. As can be seen, even the pair of curves corresponding to the $90\%$ quantile, are basically indistinguishable. In Fig. \ref{fig5} we have illustrated the case when the cosmological distortion is ignored. One can see that for $\sim 90\%$ of cases we make relative errors of $\sim 6\%$, which is comparable to the amplitude of the acoustic oscillatory features. The lower panel of Fig. \ref{fig5} presents pairs of spectra for $68\%$ and $90\%$ quantiles of the $\chi^2$ values. The inset shows the probability distribution function for the ``isotropized'' dilation scale, as given in Eq. (\ref{eq11}), compatible with the {\sc Wmap} plus HST Key project constraints. Since the values of $c_{\mathrm{isotr}}$ are quite often seen to differ from $c_{\mathrm{isotr}}=1$ by $5-10\%$, it is clear that the cosmological distortion has to be taken into account if the power spectrum is measured as accurately as given by the SDSS LRG data points in the lower panel of Fig. \ref{fig5}.      

\section{Results}
\subsection{{\sc Wmap} + HST data}
\begin{figure}
\centering
\includegraphics[width=\plotwd]{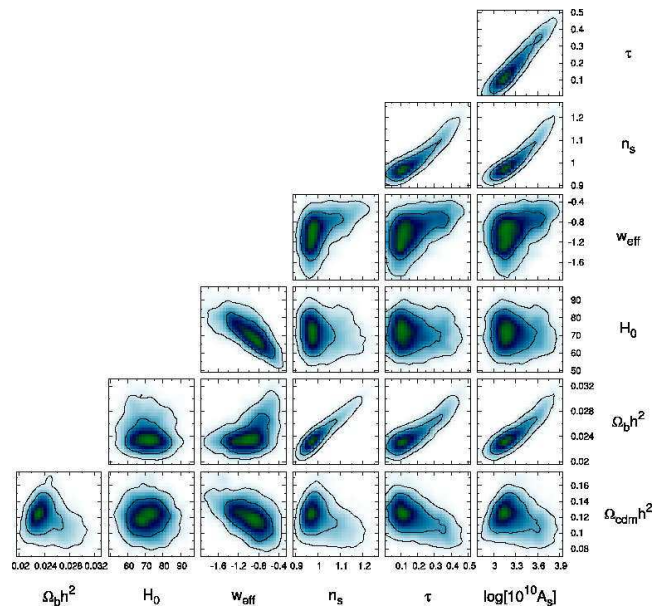}
\caption{The 2D marginalized distributions for the {\sc Wmap} + HST data.}
\label{fig6}
\end{figure}

As a starting point for several subsequent calculations we build a Markov chain using the {\sc Wmap} temperature-temperature \citep{2003ApJS..148..135H} and temperature-polarization \citep{2003ApJS..148..161K} angular spectra in combination with the constraint on the Hubble parameter from the HST Key Project \citep{2001ApJ...553...47F}. The results for the 2D marginalized distributions for all of the involved parameter pairs are shown in Fig. \ref{fig6}. Here the $68\%$ and $95\%$ credible regions are shown by solid lines. The original MCMC calculation as performed by the {\sc Cosmomc} software uses the variable $\theta$ -- the angle subtended by the sound horizon at last scattering-- in place of the more common Hubble parameter $H_0$. This leads to the better mixing of the resulting chain since $\theta$ is only weakly correlated with other variables \citep{2002PhRvD..66f3007K}. The proposal distribution for all of the MCMC calculations carried out in this paper is taken to be a multivariate Gaussian. For the current {\sc Wmap} + HST case we have used the CMB parameter covariance matrix as provided by the {\sc Cosmomc} package. All of the seven default parameters here get implicit flat priors. The marginalized distributions in Fig. \ref{fig6} are derived from a $100,000$-element Markov chain. As there is a very good proposal distribution available the chains typically equilibrate very fast and only a few hundred first elements need to be removed to eliminate the effects of the initial transients. We determine the length of this so-called burn-in period using the Gibbsit \footnote{http://www.stat.washington.edu/raftery/software.html} software \citep{rafterylewis}. The same program can also be used to estimate the length of the Markov chain required to achieve a desired accuracy for the parameter measurements. As a test one can run initially a short chain of a few thousand elements and analyze it with Gibbsit. It turned out that in the current case if we would like to achieve a $1.25\%$ accuracy at $95\%$ confidence level for the measurement of the $2.5\%$ and $97.5\%$-quantiles of the most poorly sampled parameter, we would need a chain of $\sim 25,000$ elements. Thus according to this result our $100,000$ element chain is certainly more than sufficient. Of course, all the various tools for diagnosing the convergence and for estimating the required chain length \footnote{For a lot of online material related to these issues see http://www.statslab.cam.ac.uk/$\sim$mcmc/ .} are just some more or less justified ``recipes'' that can lead to strongly incorrect results, especially in cases of poorly designed proposal distributions. Luckily, in cosmology as we have a very good knowledge about the possible parameter degeneracies, and also as the parameter spaces are relatively low dimensional, the construction of very good samplers is not too difficult. 

In the following subsections we use this {\sc Wmap} + HST chain for the very fast determination of the parameter constraints resulting from the additional measurement of the SDSS LRG acoustic scale. The same chain was also used to produce Figs. \ref{fig4} and \ref{fig5}.   

\subsection{Constraints from the measurement of the acoustic scale}\label{sect4}
\begin{figure}
\centering
\includegraphics[width=\plotwd]{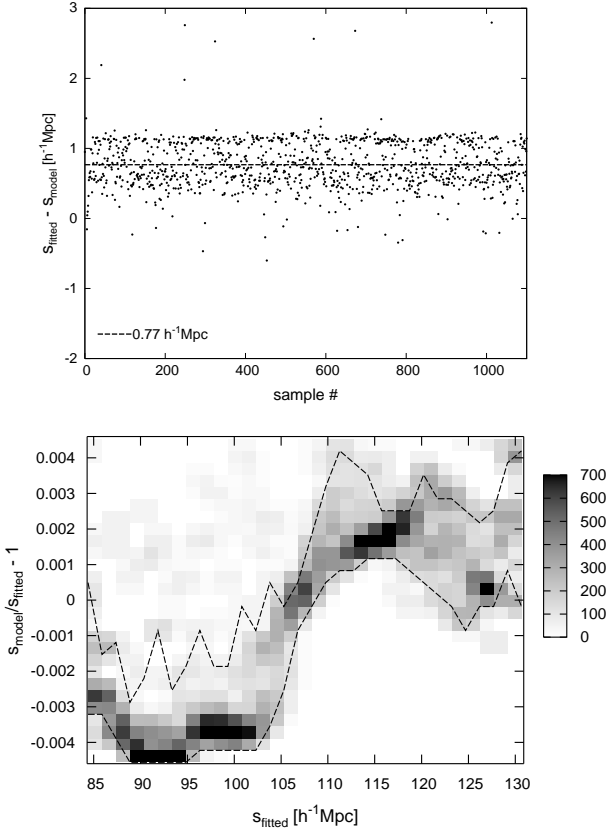}
\caption{Upper panel: Comparison of the sound horizon as determined from $\sim 1000$ model spectra, via the same fitting techniques that were used in \citet{astro-ph/0512201} to measure the SDSS LRG sound horizon, with the analytical approximation given in Eqs. (\ref{eq1_26}), (\ref{eq1_27}), (\ref{eq1_30}), (\ref{eq1_31}), (\ref{eq1_32}). The model spectra were drawn from the posterior distribution corresponding to the {\sc Wmap} + HST data. Lower panel: The density plot of the residuals after removing the average bias of $0.77\,h^{-1}\,\mathrm{Mpc}$. The solid dashed lines mark the $68\%$ credible region.}
\label{fig7}
\end{figure}

The low redshift acoustic scale as measured via the analysis of the SDSS LRG power spectrum was found to be $(105.4 \pm 2.3)\,h^{-1}\,\mathrm{Mpc}$ if adiabatic initial conditions were assumed (i.e. allowing only for the sinusoidal modulation in the spectrum), and $(103.0 \pm 7.6)\,h^{-1}\,\mathrm{Mpc}$ if this assumption was relaxed by allowing additional oscillation phase shifts \citep{astro-ph/0512201}. These measurements refer to the {\sc Wmap} best-fit cosmology \citep{2003ApJS..148..175S} which was used to analyze the SDSS LRG data. In Sect. \ref{sect3} we have described accurate transformations needed to accommodate other background cosmologies. In the following we use SH1 and SH2 to denote the sound horizon measurements of $(105.4 \pm 2.3)\,h^{-1}\,\mathrm{Mpc}$ and $(103.0 \pm 7.6)\,h^{-1}\,\mathrm{Mpc}$, respectively. 

In this section we investigate the constraints on cosmological parameters using the above given values for the sound horizon in combination with the {\sc Wmap} data. We obtain initial bounds on parameters in a numerically efficient way by applying the method of importance sampling on the earlier calculated {\sc Wmap} + HST chain. However, to be confident in the results obtained we always carry out a full MCMC calculation from scratch for each of the considered cases. As a final results we only quote the constraints on cosmological parameters obtained from the direct MCMC calculations. Importance sampling is only used as an independent check of the validity of the results. In general both methods reach to the parameter bounds that are in a good agreement. 

It is fine to use importance sampling if new constraints are not too constraining and are consistent with the earlier generated chain. Having a measurement of the acoustic scale $\widetilde{s}$ \footnote{We use tilde to denote the quantities that are ``tied to the'' fiducial cosmological model used to analyze the data.} with an error $\Delta \widetilde{s}$, importance sampling simply amounts to multiplying each original sample weight by
\begin{equation}\label{eq13}
f_i = \exp\left[-\frac{(\widetilde{s}_{\mathrm{model}_i} - \widetilde{s})^2}{2\,\Delta\widetilde{s}^2}\right]\,,
\end{equation}    
where $\widetilde{s}_{\mathrm{model}_i}$ denotes the theoretical sound horizon corresponding to the $i$-th Markov chain element. The physical size of the sound horizon $s$ at the end of the drag-epoch is is determined by the parameter combinations $\Omega_m h^2$ and $\Omega_b h^2$ i.e. physical densities of the CDM and baryonic components. Accurate fitting formulae for $s$ can be found in \citet{1996ApJ...471..542H,1998ApJ...496..605E}. We have provided these in Appendix \ref{appb} where also the transformation into different cosmological frame is described. This transformation induces an extra dependence of the sound horizon $\widetilde{s}$, as measured from the matter power spectrum, on $h$ and $w_{\mathrm{eff}}$. For more details see Appendix \ref{appb}. The dependence of $\widetilde{s}$ at redshift $z\sim 0.35$ on various parameters for spatially flat models around the best fitting {\sc Wmap} model point can be conveniently expressed as the following principal component:
\begin{equation}\label{eq14}
\left(\frac{\Omega_m h^2}{0.14}\right)^{-0.28}\left(\frac{\Omega_b h^2}{0.022}\right)^{-0.10}\left(\frac{h}{0.71}\right)^{0.94}\left(\frac{w_{\mathrm{eff}}}{-1.}\right)^{0.14} = 1 \pm \frac{\Delta \widetilde{s}}{\widetilde{s}}\,.
\end{equation} 
As probably expected, for those relatively small redshifts by far the strongest dependence is on the Hubble parameter $h$. 
To avoid any biases due to the approximate nature of the Eqs. (\ref{eq1_30}),(\ref{eq1_31}),(\ref{eq1_32}), and also due to the specific method used to measure the oscillation frequency in the SDSS LRG power spectrum, we carry out the following Monte Carlo study. We draw $\sim 1000$ samples from the {\sc Wmap} + HST chain by thinning it by a factor of $\sim 10$. For each of the parameter combinations we calculate theoretical matter spectra using {\sc Camb}. The oscillatory components of the spectra are extracted by dividing them with a ``smoothed'' approximate model spectra as given in \citet{1998ApJ...496..605E}. \footnote{The separation of the oscillatory component and the underlying smooth CDM continuum can be done very cleanly due to significantly different characteristic scales over which they change. The small residual deformations of the oscillatory part have negligible impact on the inferred oscillation period.} The resulting ``flattened'' spectra are fitted with damped sinusoidal waves \footnote{For a precise parametric form see \citet{astro-ph/0512201}.} and the sound horizon $\widetilde{s}_{\mathrm{fitted}}$ is determined via the Levenberg-Marquardt fitting technique. All the spectra are calculated at exactly the same wavenumbers as the data points given in Fig \ref{fig1}. The power spectrum covariance matrix is taken from the Appendix G of \citet{astro-ph/0512201}. For each model the sound horizon $\widetilde{s}_{\mathrm{model}}$ is calculated using Eqs. (\ref{eq1_26}), (\ref{eq1_27}), (\ref{eq1_30}), (\ref{eq1_31}), (\ref{eq1_32}). The comparison of $\widetilde{s}_{\mathrm{fitted}}$ versus $\widetilde{s}_{\mathrm{model}}$ is provided in Fig. \ref{fig7}. In the upper panel we have plotted $\widetilde{s}_{\mathrm{fitted}}-\widetilde{s}_{\mathrm{model}}$. As can be seen there is a slight tendency for the fitted values $\widetilde{s}_{\mathrm{fitted}}$ to be larger than $\widetilde{s}_{\mathrm{model}}$. After removing the constant bias of $0.77\,h^{-1}\,\mathrm{Mpc}$ the remaining fluctuations are $\lesssim 0.5 \%$, which is demonstrated in the lower panel of Fig. \ref{fig7}. This plot is an analog to the earlier density plots shown in Figs. \ref{fig2}, \ref{fig3}, \ref{fig4}, \ref{fig5}. Here the dashed lines show the region covering $68\%$ of all the cases. These were the results that apply to the case when the phase of the sinusoidal waves was not allowed to vary. If the phase is additionally allowed to change the corresponding effective bias turns out to be in the opposite direction with a value $1.8\,h^{-1}\,\mathrm{Mpc}$, instead. Thus the bias corrected values for the sound horizon used in our analysis are the following: $(104.6 \pm 2.3)\,h^{-1}\,\mathrm{Mpc}$ (SH1) and $(104.8 \pm 7.6)\,h^{-1}\,\mathrm{Mpc}$ (SH2). One might even go a step further and instead of removing only a constant offset, remove also the next order i.e. the linear component. This more accurate treatment has probably rather negligible effect on the final results, since around the measured sound horizon values of $\sim 105\,h^{-1}\,\mathrm{Mpc}$ the accuracy after removing the constant offset is already $\sim 0.2-0.3\%$, which is an order of magnitude smaller than the measurement errors of $2.3-7.6\,h^{-1}\,\mathrm{Mpc}$. 

\begin{figure}
\centering
\includegraphics[width=\plotwd]{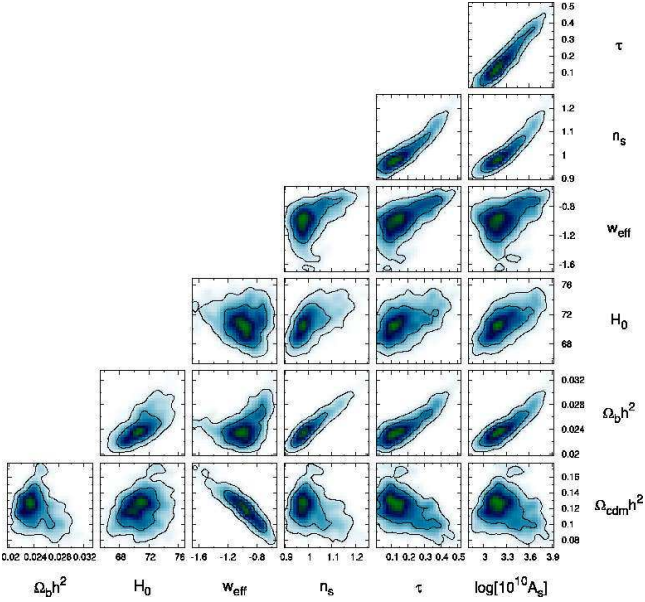}
\caption{The 2D marginalized distributions for the {\sc Wmap} data along with the constraint on the low redshift sound horizon, $\widetilde{s}=(104.6 \pm 2.3)\,h^{-1}\,\mathrm{Mpc}$ (SH1), obtained via the importance sampling of the {\sc Wmap} + HST results shown in Fig. \ref{fig6}.}
\label{fig8}
\end{figure}

\begin{figure}
\centering
\includegraphics[width=\plotwd]{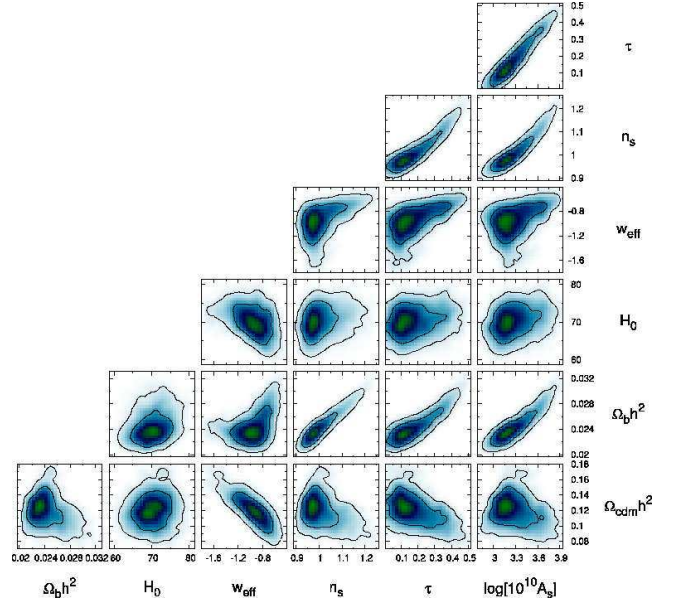}
\caption{The same as Fig. \ref{fig8}, only for the sound horizon measurement $(104.8 \pm 7.6)\,h^{-1}\,\mathrm{Mpc}$ (SH2).}
\label{fig9}
\end{figure}

\begin{figure}
\centering
\includegraphics[width=\plotwd]{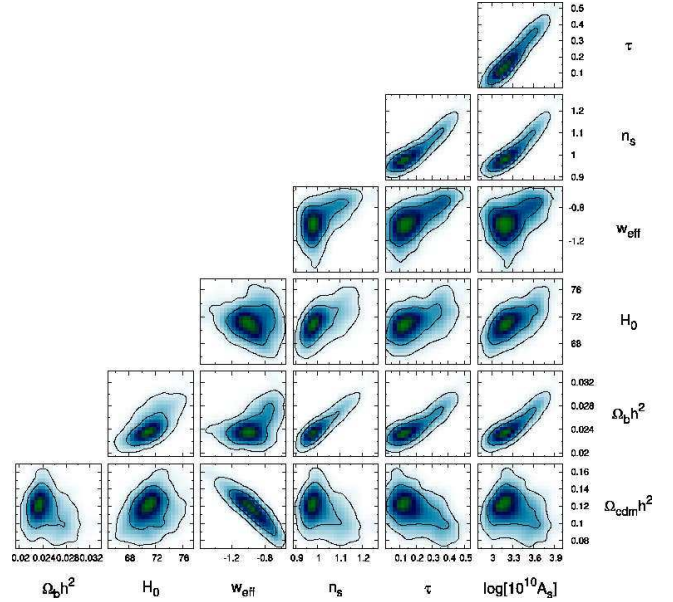}
\caption{The exact analog of Fig. \ref{fig8}, now for the full MCMC calculation.}
\label{fig10}
\end{figure}

Using this correction for the bias and the method to calculate the theoretical size of the sound horizon at the end of the drag-epoch, as presented in Appendix \ref{appb}, we can immediately perform the relevant reweighting of the {\sc Wmap} + HST chain (see Eq. (\ref{eq13})). It turns out that relatively large fraction of the {\sc Wmap} + HST chain elements ``survive'' this reweighting procedure, justifying the use of the importance sampling method. In particular, for the SH2 case we are left with $\sim 36,000$, and for the SH1 $\sim 12,000$ samples. The results of this calculation in the form of the 2D marginalized distributions is presented in Figs. \ref{fig8} and \ref{fig9}. Here Fig. \ref{fig8}/\ref{fig9} corresponds to the SH1/SH2 case. In comparison to the analogous Fig. \ref{fig6} the most dramatic changes are for $H_0$ and $w_{\mathrm{eff}}$, whereas the rest of the parameters stay essentially the same. The HST constraint for the initial {\sc Wmap} chain was just implemented in order not to loose too many samples in current importance sampling calculations. As can be seen, the new constraints on $H_0$ are significantly stronger than the one provided by the HST. In Fig. \ref{fig8} due to somewhat lower number of samples ($\sim 12,000$) the contours start to become more noisy. Earlier we estimated that the measurement of the $2.5\%$ and $97.5\%$ quantiles with an accuracy of $1.25\%$ at $95\%$ CL requires $\sim 25,000$ samples. So is this $12,000$ enough for the parameter estimation purposes? To test that we have also performed a full MCMC calculation from scratch (with $100,000$ samples) using {\sc Wmap} data along with a sound horizon measurement SH1. The results of this calculation are shown in Fig. \ref{fig10}. The contours in Fig. \ref{fig8} although being noisier are very similar to the ones in Fig. \ref{fig10}. In fact the corresponding 1D distributions are practically indistinguishable. This shows that the initial use of importance sampling was indeed justified.          

\subsection{Constraints from the full power spectrum}\label{sect5}

\begin{figure}
\centering
\includegraphics[width=\plotwd]{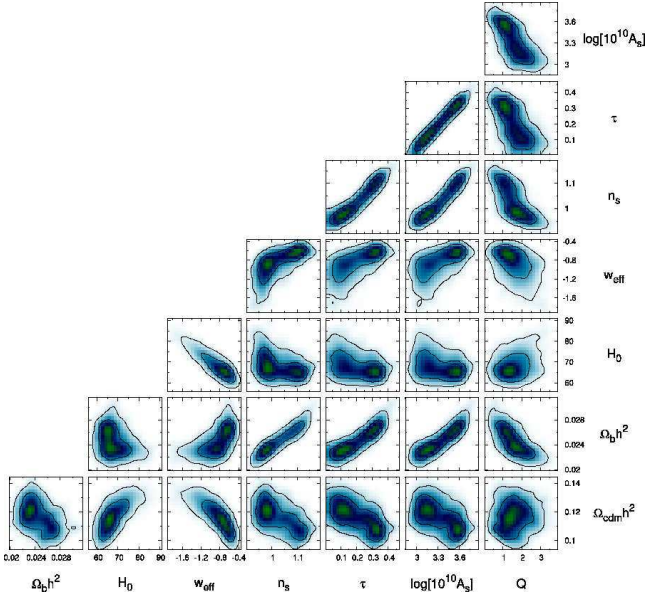}
\caption{The 2D marginalized distributions from the {\sc Wmap} + SDSS LRG full power spectrum MCMC calculation.}
\label{fig11}
\end{figure}

Using the {\sc Wmap} data and the SDSS LRG power spectrum as shown in Fig. \ref{fig1} along with the power spectrum transformation and an additional new parameter $Q$, as described in Sect. \ref{sect3}, we build a $100,000$ element Markov chain in the $8$-dimensional parameter space. The resulting 2D parameter distribution functions are shown in Fig. \ref{fig11}. Here we see that in several cases distributions start to become doubly-peaked. Also the constraints on $H_0$ and $w_{\mathrm{eff}}$ are weaker than the ones obtained in the previous subsection. On the other hand, now a rather strong constraint has been obtained for $\Omega_{cdm}h^2$. Even stronger constraint (not shown in the figure) is obtained for $\Omega_mh$-- the shape parameter $\Gamma$. This just illustrates the the well-known fact that the shape of the matter power spectrum is most sensitive to $\Gamma$. The new parameter $Q$, describing the deformation of the linear spectrum to the evolved redshift-space galaxy power spectrum, is seen to be significantly degenerate with several parameters e.g. $\Omega_bh^2$, $n_s$, $\tau$, $A_s$. On the other hand it does not interfere too strongly with $H_0$.

It might seem strange that using the full data we obtain weaker constraints on $H_0$ and $w_{\mathrm{eff}}$. But after all, we should not be too surprised, since our understanding of how the linear spectrum is deformed to the evolved redshift-space power spectrum is rather limited. Here we were introducing an additional parameter $Q$, which starts to interfere with the rest of the parameter estimation. Also one should remind that maximum likelihood is the global fitting technique i.e. it is not very sensitive to specific features in the data. On the other hand, modeling of the oscillatory component of the spectrum does not call for any extra parameters. Also the underlying physics is much better understood. In fact, the observable low redshift acoustic scale is determined by four parameters only: $\Omega_mh^2$, $\Omega_bh^2$, $w_{\mathrm{eff}}$ and $h$. The optimal data analysis of course should incorporate both components: (i) general shape of the spectrum i.e. low frequency components, and (ii) oscillatory part, with appropriate weightings. It is clear that in the current ``full spectrum'' maximum likelihood analysis the acoustic features are weighted too weakly.       

\subsection{One dimensional distributions}

\begin{figure}
\centering
\includegraphics[width=\plotwd]{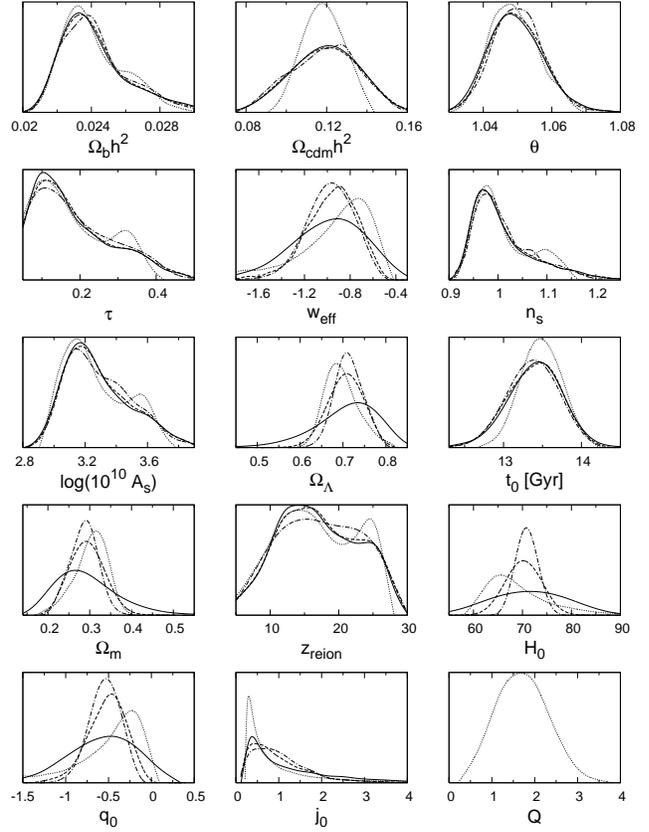}
\caption{The 1D posterior distributions for several cosmological parameters. Solid, dash-dotted, dashed, and dotted lines correspond to the {\sc Wmap} + HST, {\sc Wmap} + SDSS LRG SH1, {\sc Wmap} + SDSS LRG SH2, and {\sc Wmap} + SDSS LRG full power spectrum cases, respectively. The compact summary of these results can be found in Table \ref{tab1}.}
\label{fig12}
\end{figure}

To compare the measurements of the parameters in a more clear fashion we provide in Fig. \ref{fig12} several 1D marginalized distributions. The $68\%$ and $95\%$ credible regions along with the medians of these distributions are provided in Table \ref{tab1}. Here the parameters $\Omega_bh^2$, $\Omega_{cdm}h^2$, $\theta$, $\tau$, $w_{\mathrm{eff}}$, $n_s$, $A_s$ and $Q$ (the last in case of the full spectrum analysis only) are primary parameters as used in the MCMC calculations. All the rest: $\Omega_{\Lambda}$, $t_0$, $\Omega_m$, $z_{\mathrm{reion}}$, $H_0$, $q_0$, $j_0$ are derived from these. Here $t_0$ is the age of the Universe, $q_0$ the deceleration parameter and $j_0$ the so-called jerk (see e.g. \citealt{2005ASPC..339...27B}) at $z=0$. The deceleration parameter $q_0$ and jerk $j_0$ are introduced as usual via the Taylor expansion of the scale factor:
\begin{eqnarray}
a(t)=a_0\Big[1+H_0(t-t_0)&-&\frac{1}{2}q_0H_0^2(t-t_0)^2 + \Big.\nonumber\\
\Big.&+&\frac{1}{6}j_0H_0^3(t-t_0)^3+\ldots\Big]\,.
\end{eqnarray}
From Fig. \ref{fig12} we can see that many parameters stay essentially the same as determined by {\sc Wmap} + HST data. On the other hand, a new precise measurement of $H_0$, thanks to the measurement of the low redshift sound horizon along with strong constraints on $\Omega_bh^2$ and $\Omega_{cdm}h^2$ from the CMB data, helps to determine $\Omega_m$ (as well as $\Omega_b$ and $\Omega_{cdm}$ separately) rather precisely. The same applies to the case of the full spectrum analysis, which provides us with a good estimate for the shape parameter $\Gamma=\Omega_mh$. In both cases also the constraint on $w_{\mathrm{eff}}$ is significantly improved. New improved limits on $\Omega_m$ and $w_{\mathrm{eff}}$ immediately transform to better constraints on $q_0$ and $j_0$ (see Appendix \ref{appc}). For the ``vanilla'' $\Lambda$CDM model with $w=-1$ the jerk parameter $j_0=1$. We can see that at the moment jerk is still rather poorly constrained. Using only the observational data whose nature is very well understood, namely the CMB power spectra along with the low redshift sound horizon measurement, we get very strong support for the accelerating Universe (i.e. $q_0 < 0$). The values $q_0 > 0$ are ruled out by $1.4\sigma$, $2.9\sigma$ and $5.5\sigma$ in case of the {\sc Wmap} + HST, {\sc Wmap} + SDSS LRG SH2 and {\sc Wmap} + SDSS LRG SH1, respectively. \footnote{We can perform this analysis of the far away tails of the distributions since the {\sc Wmap} + HST chain we start with contains enough samples with $q_0>0$ (see Fig. \ref{fig12}).} Of course, one has to remind that until now the analysis assumed flat spatial sections.      

\begin{table*}
\caption{Various quantiles of the 1D distributions shown in Fig. \ref{fig12}. The first group of parameters are the primary ones used in the MCMC calculations, the second group represents various derived quantities, and the last shows the parameters held fixed due to our prior assumptions. The last row of the table also gives the total number of free parameters, excluding the bias parameter that was marginalized out analytically, for all of the investigated cases. Also shown are the $\chi^2$-values for the best-fitting model and the effective number of degrees of freedom involved.}
{\footnotesize
\vspace{0.5cm}
\label{tab1}
\begin{tabular}{|l|l|l|l|l|l|l|l|l|l|l|l|l|}
\hline
 & \multicolumn{3}{c|}{WMAP + HST} & \multicolumn{3}{c|}{WMAP + SDSS LRG} & \multicolumn{3}{c|}{WMAP + SDSS LRG} & \multicolumn{3}{c|}{WMAP + SDSS LRG}\\
Parameter & \multicolumn{3}{c|}{} & \multicolumn{3}{c|}{sound horizon; adiab. (SH1)} & \multicolumn{3}{c|}{sound horizon (SH2)} & \multicolumn{3}{c|}{full spectrum}\\
\cline{2-13}
 & Median & 68\% & 95\% & Median & 68\% & 95\% & Median & 68\% & 95\% & Median & 68\% & 95\% \\
\hline
$\Omega_bh^2$ & $0.0238$ & ${}^{+0.0026}_{-0.0014}$ & ${}^{+0.0057}_{-0.0025}$ &$0.0239$ & ${}^{+0.0028}_{-0.0014}$ & ${}^{+0.0062}_{-0.0027}$ & $0.0238$ & ${}^{+0.0024}_{-0.0014}$ & ${}^{+0.061}_{-0.025}$ & $0.0237$ & ${}^{+0.0025}_{-0.0013}$ & ${}^{+0.0043}_{-0.0023}$\\
$\Omega_{cdm}h^2$ & $0.120$ & ${}^{+0.017}_{-0.018}$ & ${}^{+0.031}_{-0.033}$ & $0.119$ & ${}^{+0.018}_{-0.018}$ & ${}^{+0.033}_{-0.035}$ & $0.118$ & ${}^{+0.018}_{-0.017}$ & ${}^{+0.036}_{-0.033}$ & $0.118$ & ${}^{+0.011}_{-0.010}$ & ${}^{+0.020}_{-0.018}$\\
$\theta$ & $1.0489$ & ${}^{+0.0085}_{-0.0073}$ & ${}^{+0.0179}_{-0.0138}$ & $1.0496$ & ${}^{+0.0088}_{-0.0081}$ & ${}^{+0.0182}_{-0.0149}$ & $1.0488$ & ${}^{+0.0083}_{-0.0070}$ & ${}^{+0.0178}_{-0.0131}$ & $1.0486$ & ${}^{+0.0075}_{-0.0065}$ & ${}^{+0.0146}_{-0.0119}$\\
$\tau$ & $0.153$ & ${}^{+0.044}_{-0.153}$ & ${}^{+0.231}_{-0.153}$ & $0.160$ & ${}^{+0.047}_{-0.160}$ & ${}^{+0.238}_{-0.160}$ & $0.161$ & ${}^{+0.043}_{-0.161}$ & ${}^{+0.239}_{-0.161}$ & $0.152$ & ${}^{+0.046}_{-0.152}$ & ${}^{+0.195}_{-0.152}$\\
$w_{\mathrm{eff}}$ & $-0.97$ & ${}^{+0.30}_{-0.36}$ & ${}^{+0.54}_{-0.73}$ & $-0.96$ & ${}^{+0.18}_{-0.22}$ & ${}^{+0.32}_{-0.45}$ & $-0.95$ & ${}^{+0.20}_{-0.23}$ & ${}^{+0.34}_{-0.53}$ & $-0.86$ & ${}^{+0.21}_{-0.40}$ & ${}^{+0.34}_{-0.92}$\\
$n_s$ & $0.991$ & ${}^{+0.084}_{-0.037}$ & ${}^{+0.183}_{-0.061}$ & $0.993$ & ${}^{+0.093}_{-0.037}$ & ${}^{+0.195}_{-0.063}$ & $0.992$ & ${}^{+0.083}_{-0.036}$ & ${}^{+0.190}_{-0.061}$ & $0.993$ & ${}^{+0.092}_{-0.036}$ & ${}^{+0.141}_{-0.060}$\\
$\log(10^{10}A_s)$ & $3.24$ & ${}^{+0.28}_{-0.16}$ & ${}^{+0.53}_{-0.28}$ & $3.26$ & ${}^{+0.31}_{-0.18}$ & ${}^{+0.53}_{-0.32}$ & $3.24$ & ${}^{+0.28}_{-0.17}$ & ${}^{+0.54}_{-0.29}$ & $3.22$ & ${}^{+0.30}_{-0.16}$ & ${}^{+0.45}_{-0.24}$\\
$q$ & --- & --- & --- & --- & --- & --- & --- & --- & --- & $1.71$ & ${}^{+0.63}_{-0.60}$ & ${}^{+1.25}_{-1.11}$\\
\hline
$H_0 [\mathrm{km/s/Mpc}]$ & $71.4$ & ${}^{+8.0}_{-8.2}$ & ${}^{+14.9}_{-15.5}$ & $70.8$ & ${}^{+2.1}_{-2.0}$ & ${}^{+4.4}_{-4.0}$ & $70.5$ & ${}^{+3.8}_{-3.7}$ & ${}^{+7.8}_{-7.3}$ & $67.6$ & ${}^{+7.7}_{-4.3}$ & ${}^{+17.9}_{-7.2}$\\
$q_0$ & $-0.54$ & ${}^{+0.38}_{-0.45}$ & ${}^{+0.68}_{-0.89}$ & $-0.53$ & ${}^{+0.16}_{-0.18}$ & ${}^{+0.29}_{-0.37}$ & $-0.51$ & ${}^{+0.20}_{-0.24}$ & ${}^{+0.35}_{-0.51}$ & $-0.38$ & ${}^{+0.24}_{-0.50}$ & ${}^{+0.38}_{-1.22}$\\
$j_0$ & $0.91$ & ${}^{+1.55}_{-0.54}$ & ${}^{+4.11}_{-0.69}$ & $0.87$ & ${}^{+0.76}_{-0.45}$ & ${}^{+1.82}_{-0.67}$ & $0.85$ & ${}^{+0.83}_{-0.47}$ & ${}^{+2.31}_{-0.65}$ & $0.63$ & ${}^{+1.43}_{-0.33}$ & ${}^{+5.31}_{-0.39}$\\
$\Omega_{\Lambda}$ & $0.717$ & ${}^{+0.063}_{-0.083}$ & ${}^{+0.108}_{-0.189}$ & $0.715$ & ${}^{+0.031}_{-0.032}$ & ${}^{+0.062}_{-0.059}$ & $0.712$ & ${}^{+0.040}_{-0.042}$ & ${}^{+0.076}_{-0.090}$ & $0.693$ & ${}^{+0.045}_{-0.034}$ & ${}^{+0.096}_{-0.064}$\\
$\Omega_m$ & $0.283$ & ${}^{+0.083}_{-0.063}$ & ${}^{+0.189}_{-0.108}$ & $0.285$ & ${}^{+0.032}_{-0.030}$ & ${}^{+0.059}_{-0.062}$ & $0.288$ & ${}^{+0.042}_{-0.040}$ & ${}^{+0.089}_{-0.077}$ & $0.307$ & ${}^{+0.034}_{-0.045}$ & ${}^{+0.064}_{-0.096}$\\
$t_0$ & $13.42$ & ${}^{+0.31}_{-0.36}$ & ${}^{+0.64}_{-0.76}$ & $13.38$ & ${}^{+0.34}_{-0.40}$ & ${}^{+0.65}_{-0.84}$ & $13.43$ & ${}^{+0.32}_{-0.39}$ & ${}^{+0.60}_{-0.87}$ & $13.48$ & ${}^{+0.27}_{-0.26}$ & ${}^{+0.53}_{-0.52}$\\
$z_{\mathrm{reion}}$ & $16.5$ & ${}^{+7.3}_{-5.5}$ & ${}^{+10.9}_{-10.3}$ & $16.9$ & ${}^{+7.8}_{-5.7}$ & ${}^{+10.6}_{-10.7}$ & $16.8$ & ${}^{+7.2}_{-5.9}$ & ${}^{+11.0}_{-10.5}$ & $16.4$ & ${}^{+7.7}_{-5.9}$ & ${}^{+10.0}_{-10.6}$\\
$\Gamma \equiv \Omega_mh$ & $0.202$ & ${}^{+0.034}_{-0.031}$ & ${}^{+0.073}_{-0.058}$ & $0.202$ & ${}^{+0.024}_{-0.023}$ & ${}^{+0.044}_{-0.044}$ & $0.202$ & ${}^{+0.026}_{-0.024}$ & ${}^{+0.050}_{-0.046}$ & $0.207$ & ${}^{+0.011}_{-0.012}$ & ${}^{+0.023}_{-0.027}$\\
\hline
$\Omega_k$ & $0$ & $0$ & $0$ & $0 $& $0$ & $0$ & $0$ & $0$ & $0$ & $0$ & $0$ & $0$\\
$\Sigma m_{\nu} [\mathrm{eV}]$ & $0$ & $0$ & $0$ & $0 $& $0$ & $0$ & $0$ & $0$ & $0$ & $0$ & $0$ & $0$\\
$A_T$ & $0$ & $0$ & $0$ & $0 $& $0$ & $0$ & $0$ & $0$ & $0$ & $0$ & $0$ & $0$\\
$\frac{{\mathrm d}\ln n_s}{{\mathrm d}\ln k}$ & $0$ & $0$ & $0$ & $0$ & $0$ & $0$ & $0$ & $0$ & $0$ & $0$ & $0$ & $0$\\
\hline
\multicolumn{1}{|c|}{$\chi^2/\mathrm{dof}$} & \multicolumn{3}{c|}{$1429/1404$} & \multicolumn{3}{c|}{$1429/1404$} & \multicolumn{3}{c|}{$1429/1404$} & \multicolumn{3}{c|}{$1439/1417$}\\
\hline
\multicolumn{1}{c|}{} & \multicolumn{3}{c|}{7 free parameters} & \multicolumn{3}{c|}{7 free parameters} & \multicolumn{3}{c|}{7 free parameters} & \multicolumn{3}{c|}{8 free parameters}\\
\cline{2-13}
\end{tabular}
}
\end{table*}

\subsection{Most interesting constraints}
\begin{figure}
\centering
\includegraphics[width=\plotwd]{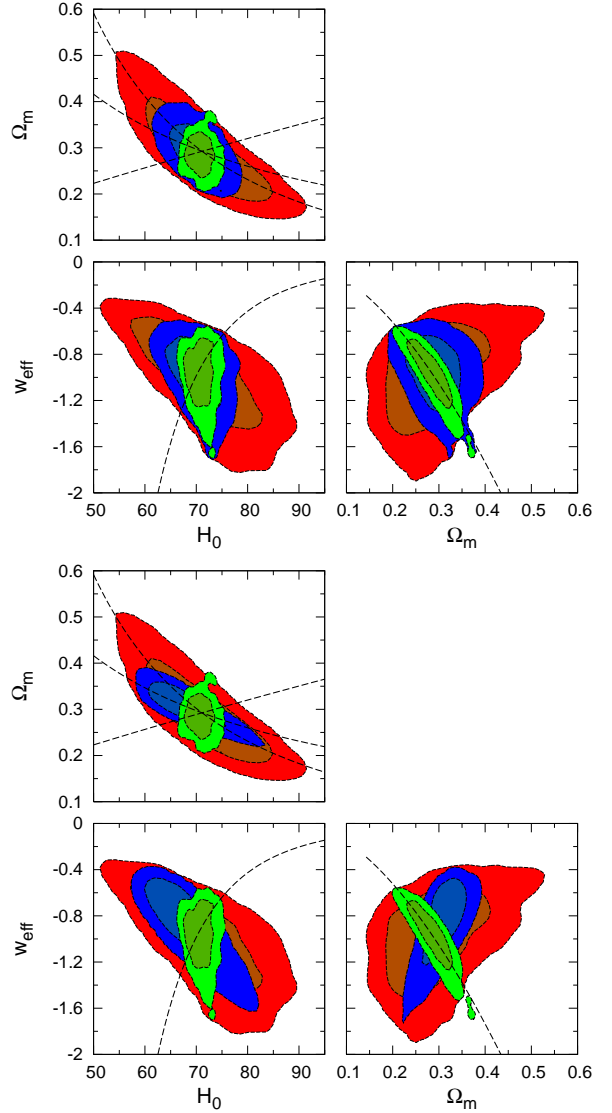}
\caption{The comparison of constraints on $H_0$, $\Omega_m$ and $w_{\mathrm{eff}}$. In all panels the largest error contours correspond to the {\sc Wmap} + HST, while the tightest to the {\sc Wmap} + SDSS LRG SH1 case. The upper group of panels shows additionally the constraints for the {\sc Wmap} + SDSS LRG SH2 case, whereas the lower group provides extra limits from the full spectrum + {\sc Wmap} analysis. The dashed lines in all the panels show the principal component from Eq. (\ref{eq14}). The additional lines in $\Omega_m-H_0$ plane provide the directions $\Omega_mh^2=\mathrm{const}$ and $\Gamma\equiv\Omega_mh=\mathrm{const}$.}
\label{fig13}
\end{figure}

\begin{figure}
\centering
\includegraphics[width=\plotwd]{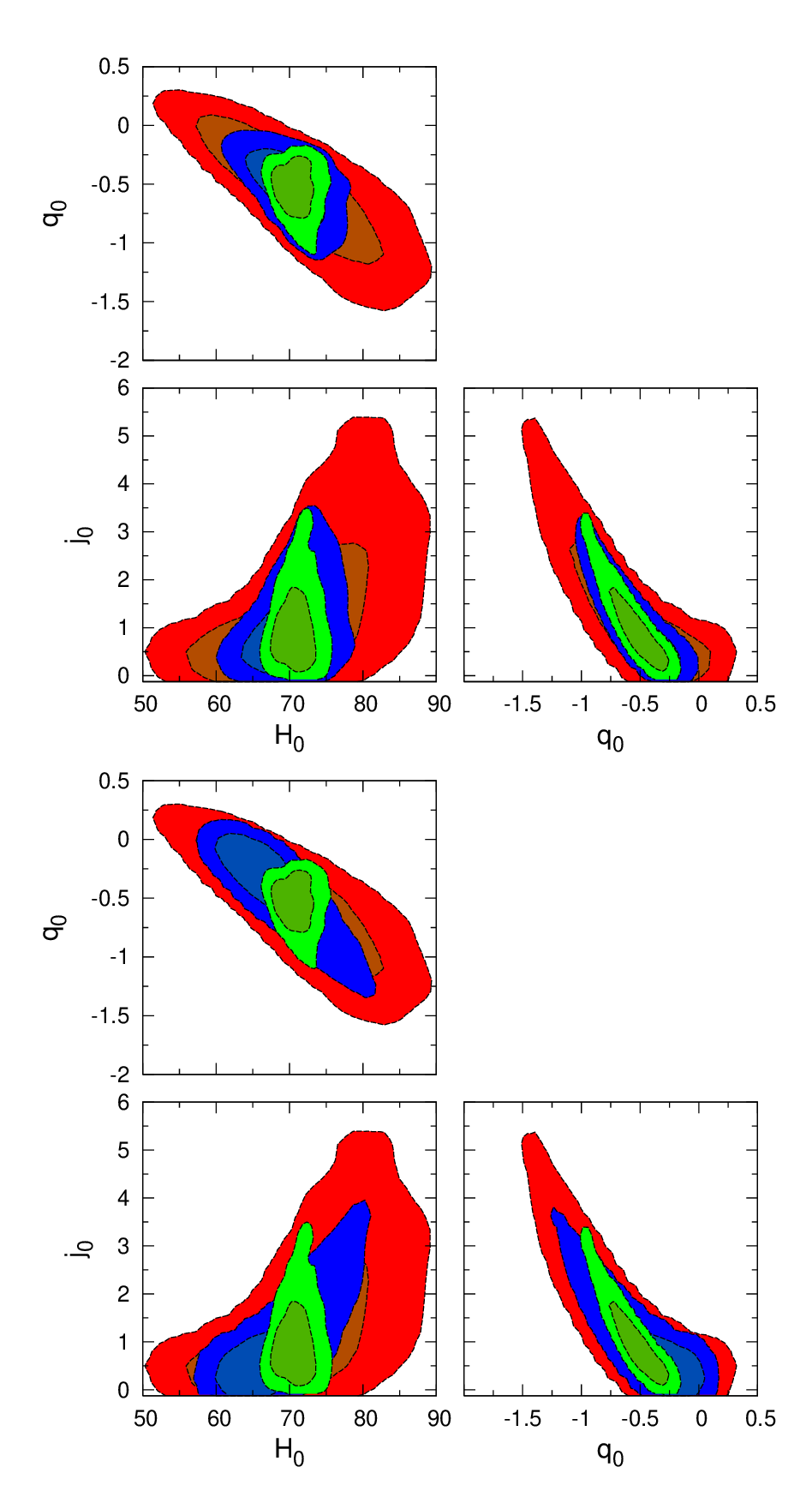}
\caption{The version of Fig. \ref{fig13} with the parameter trio ($H_0$, $\Omega_m$, $w_{\mathrm{eff}}$) replaced by ($H_0$, $q_0$, $j_0$).}
\label{fig14}
\end{figure}
\begin{figure}
\centering
\includegraphics[width=\plotwd]{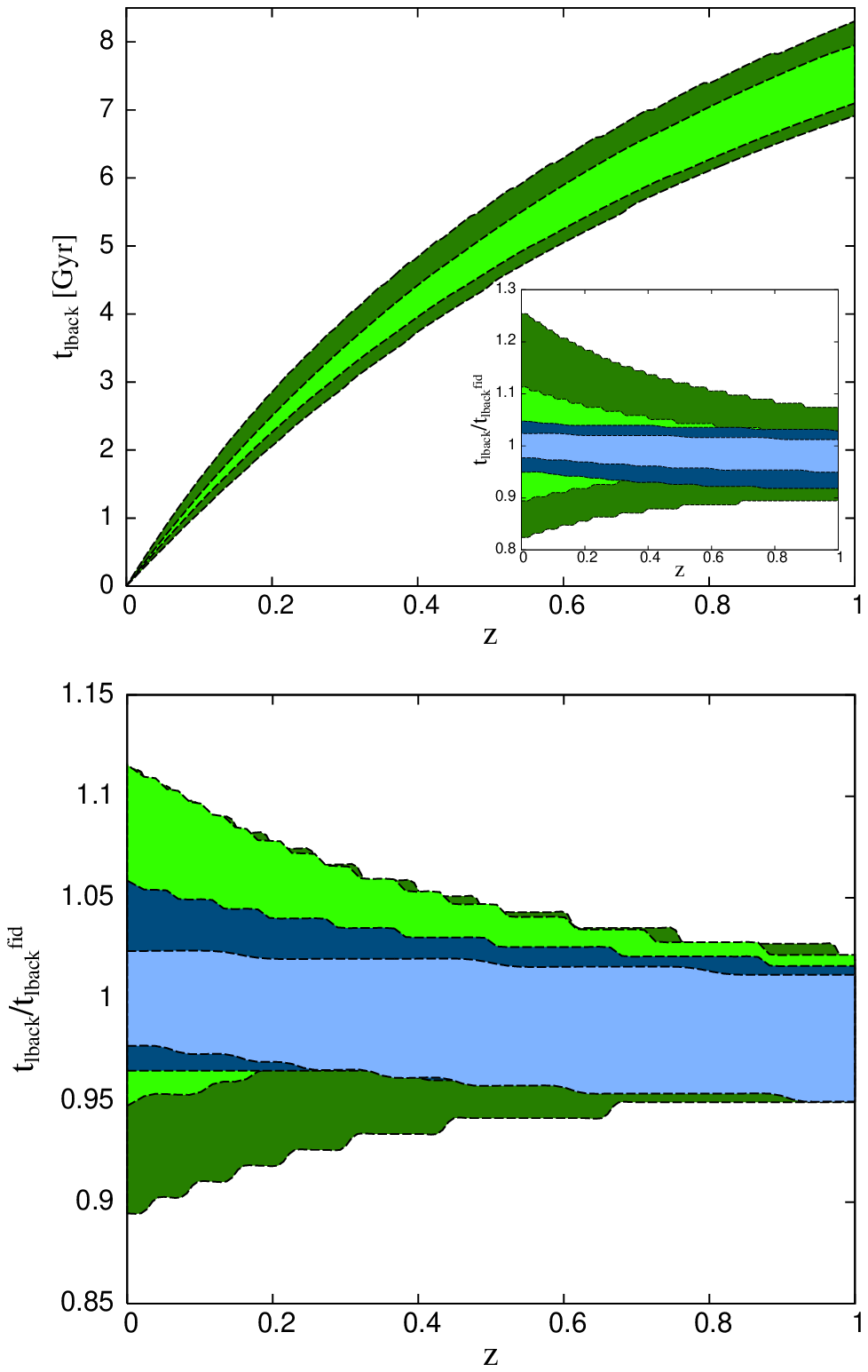}
\caption{Constraints on the low redshift expansion law. Upper panel: $2\sigma$ credible regions of the look-back time as a function of redshift for the {\sc Wmap} + HST and {\sc Wmap} + SDSS LRG SH1 cases. The inset displays these regions after dividing by the look-back time corresponding to the best-fit {\sc Wmap} cosmology. Here in addition to the $2\sigma$ contours also $1\sigma$ regions are given. Lower panel: Analog of the inset in the upper panel. Here we have given the $1\sigma$ regions as a function of redshift for (starting from the bottommost) the {\sc Wmap} + HST, {\sc Wmap} + SDSS LRG full spectrum, {\sc Wmap} + SDSS LRG SH2 and {\sc Wmap} + SDSS LRG SH1 cases.}
\label{fig15}
\end{figure}

We have shown that by adding the SDSS LRG clustering data to the {\sc Wmap} results we can get significantly tighter constraints on $H_0$, $\Omega_m$ and $w_{\mathrm{eff}}$ (or $q_0$ and $j_0$) than from the {\sc Wmap} + HST analysis alone. The comparison of the obtained limits on parameters $H_0$, $\Omega_m$ and $w_{\mathrm{eff}}$ is provided in Fig. \ref{fig13}. The largest error contours in both upper and lower group of panels correspond to {\sc Wmap} + HST, while the tightest to the {\sc Wmap} + SDSS LRG SH1 case. In the upper group of panels we have additionally given the constraints for the {\sc Wmap} + SDSS LRG SH2 case, whereas the lower group provides extra limits from the full spectrum + {\sc Wmap} analysis. In addition, in each of the panels we have given the degeneracy lines corresponding to the principal component given in Eq. (\ref{eq14}). \footnote{The analog of Eq. (\ref{eq14}) valid for the non-flat cases is given in Appendix \ref{appb}.} In $\Omega_m-H_0$ plane we have additionally plotted the lines corresponding to $\Omega_mh^2=\mathrm{const}$ and $\Gamma\equiv\Omega_mh=\mathrm{const}$. These are the combinations well determined by the CMB data and by the general shape of the matter power spectrum, respectively. As is evident from Fig. \ref{fig13}, the principal direction of the low redshift sound horizon constraint is always almost perpendicular to the corresponding {\sc Wmap} + HST error contours, demonstrating the high level of complementarity of this new measurement. For the spatially flat models with constant dark energy equation of state parameter there exists a unique relation between parameter pairs ($\Omega_m$, $w_{\mathrm{eff}}$) and ($q_0$, $j_0$) (see Appendix \ref{appc}). Fig. \ref{fig14} presents similar plots to Fig. \ref{fig13}, now only for the parameter triad ($H_0$, $q_0$, $j_0$) instead. The parameters shown in Figs. \ref{fig13} and \ref{fig14} are the ones that determine the low redshift expansion law.
Introducing the look-back time $t_{\mathrm{lb}}=t_0-t$, where $t_0$ is the age of the Universe at $z=0$, one can find for the redshift:
\begin{equation}   
z \simeq H_0t_{\mathrm{lb}} + \left(1+\frac{q_0}{2}\right)H_0^2t_{\mathrm{lb}}^2 + \left(1+q_0+\frac{j_0}{6}\right)H_0^3t_{\mathrm{lb}}^3 + \ldots
\end{equation}   
The precise calculation for the look-back time as a function of redshift is shown in Fig. \ref{fig15}. Here the upper panel shows the $2\sigma$ regions corresponding to the {\sc Wmap} + HST and {\sc Wmap} + SDSS LRG SH1, respectively. The inset in the upper panel displays these regions after dividing by the look-back time corresponding to the best-fit {\sc Wmap} cosmology. Here in addition to the $2\sigma$ contours also $1\sigma$ regions are given. It is evident that the low redshift sound horizon measurement has helped to determine the recent expansion history of the Universe with much greater accuracy than available from the {\sc Wmap} + HST data alone. Of course, this is largely due to the much tighter constraint obtained for the Hubble parameter. The lower panel in Fig. \ref{fig15} shows a similar plot than the inset in the upper panel. Here we have given only the $1\sigma$ regions as a function of redshift for (starting from the bottommost) the {\sc Wmap} + HST, {\sc Wmap} + SDSS LRG full spectrum, {\sc Wmap} + SDSS LRG SH2 and {\sc Wmap} + SDSS LRG SH1 cases.     

\section{Extended analysis}

\begin{table*}
\centering
\caption{Constraints on selected parameters from an extended analysis. Also shown are the $\chi^2$-values for the best-fitting model and the effective number of degrees of freedom involved.}
{\scriptsize
\vspace{0.5cm}
\label{tab2}
\begin{tabular}{|l|l|l|l|l|l|l|l|l|l|}
\cline{2-10}
\multicolumn{1}{c|}{} & \multicolumn{9}{c|}{7 free parameters: $\Omega_bh^2$, $\Omega_{cdm}h^2$, $\theta$, $\tau$, $n_s$, $A_s$, $\Omega_k$}\\
\cline{2-10}
\multicolumn{1}{c|}{} & \multicolumn{9}{c|}{fixed parameters: $w_{\mathrm{eff}}=-1$, $\Sigma m_{\nu}=0$, $A_T=0$, $\frac{{\mathrm d}\ln n_s}{{\mathrm d}\ln k}=0$}\\
\hline
 & \multicolumn{3}{c|}{WMAP + HST} & \multicolumn{3}{c|}{WMAP + SDSS LRG} & \multicolumn{3}{c|}{WMAP + SDSS LRG}\\
Parameter & \multicolumn{3}{c|}{} & \multicolumn{3}{c|}{sound horizon; adiab. (SH1)} & \multicolumn{3}{c|}{sound horizon (SH2)}\\
\cline{2-10}
 & Median & 68\% & 95\% & Median & 68\% & 95\% & Median & 68\% & 95\%\\
\hline
$H_0 [\mathrm{km/s/Mpc}]$ & $67.8$ & ${}^{+7.5}_{-7.9}$ & ${}^{+14.7}_{-15.1}$ & $66.8$ & ${}^{+3.6}_{-3.8}$ & ${}^{+6.8}_{-6.9}$ & $65.9$ & ${}^{+5.3}_{-4.8}$ & ${}^{+10.4}_{-9.9}$\\
$\Omega_m$ & $0.261$ & ${}^{+0.107}_{-0.068}$ & ${}^{+0.246}_{-0.114}$ & $0.261$ & ${}^{+0.027}_{-0.028}$ & ${}^{+0.048}_{-0.053}$ & $0.266$ & ${}^{+0.049}_{-0.041}$ & ${}^{+0.101}_{-0.074}$\\
$\Omega_{\Lambda}$ & $0.767$ & ${}^{+0.074}_{-0.110}$ & ${}^{+0.119}_{-0.226}$ & $0.777$ & ${}^{+0.049}_{-0.056}$ & ${}^{+0.090}_{-0.092}$ & $0.768$ & ${}^{+0.061}_{-0.065}$ & ${}^{+0.102}_{-0.120}$\\
$\Omega_k$ & $-0.029$ & ${}^{+0.023}_{-0.025}$ & ${}^{+0.041}_{-0.060}$ & $-0.037$& ${}^{+0.025}_{-0.024}$ & ${}^{+0.047}_{-0.042}$ & $-0.036$ & ${}^{+0.026}_{-0.024}$ & ${}^{+0.047}_{-0.053}$\\
$q_0$ & $-0.64$ & ${}^{+0.16}_{-0.10}$ & ${}^{+0.36}_{-0.17}$ & $-0.647$ & ${}^{+0.069}_{-0.063}$ & ${}^{+0.114}_{-0.116}$ & $-0.635$ & ${}^{+0.088}_{-0.082}$ & ${}^{+0.170}_{-0.137}$\\
$\Gamma \equiv \Omega_mh$ & $0.179$ & ${}^{+0.051}_{-0.038}$ & ${}^{+0.098}_{-0.065}$ & $0.174$ & ${}^{+0.025}_{-0.026}$ & ${}^{+0.047}_{-0.048}$ & $0.177$ & ${}^{+0.030}_{-0.031}$ & ${}^{+0.057}_{-0.052}$\\
\hline
\multicolumn{1}{|c|}{$\chi^2/\mathrm{dof}$} & \multicolumn{3}{c|}{$1429/1404$} & \multicolumn{3}{c|}{$1429/1404$} & \multicolumn{3}{c|}{$1429/1404$}\\
\hline
\end{tabular}
\\[0.5cm]
\begin{tabular}{|l|l|l|l|l|l|l|l|l|l|}
\cline{2-10}
\multicolumn{1}{c|}{} & \multicolumn{9}{c|}{8 free parameters: $\Omega_bh^2$, $\Omega_{cdm}h^2$, $\theta$, $\tau$, $n_s$, $A_s$, $w_{\mathrm{eff}}$, $\Omega_k$}\\
\cline{2-10}
\multicolumn{1}{c|}{} & \multicolumn{9}{c|}{fixed parameters: $\Sigma m_{\nu}=0$, $A_T=0$, $\frac{{\mathrm d}\ln n_s}{{\mathrm d}\ln k}=0$}\\
\hline
 & \multicolumn{3}{c|}{WMAP + HST} & \multicolumn{3}{c|}{WMAP + SDSS LRG} & \multicolumn{3}{c|}{WMAP + SDSS LRG}\\
Parameter & \multicolumn{3}{c|}{} & \multicolumn{3}{c|}{sound horizon; adiab. (SH1)} & \multicolumn{3}{c|}{sound horizon (SH2)}\\
\cline{2-10}
 & Median & 68\% & 95\% & Median & 68\% & 95\% & Median & 68\% & 95\%\\
\hline
$H_0 [\mathrm{km/s/Mpc}]$ & $63.3$ & ${}^{+9.7}_{-13.6}$ & ${}^{+18.4}_{-15.7}$ & $65.6$ & ${}^{+4.0}_{-4.4}$ & ${}^{+8.1}_{-9.5}$ & $64.8$ & ${}^{+5.4}_{-5.9}$ & ${}^{+11.4}_{-11.4}$\\
$\Omega_m$ & $0.32$ & ${}^{+0.39}_{-0.10}$ & ${}^{+0.47}_{-0.16}$ & $0.292$ & ${}^{+0.043}_{-0.055}$ & ${}^{+0.073}_{-0.092}$ & $0.291$ & ${}^{+0.060}_{-0.053}$ & ${}^{+0.118}_{-0.095}$\\
$\Omega_{\Lambda}$ & $0.71$ & ${}^{+0.11}_{-0.46}$ & ${}^{+0.18}_{-0.49}$ & $0.744$ & ${}^{+0.072}_{-0.051}$ & ${}^{+0.117}_{-0.079}$ & $0.746$ & ${}^{+0.063}_{-0.061}$ & ${}^{+0.117}_{-0.115}$\\
$w_{\mathrm{eff}}$ & $-0.94$ & ${}^{+0.87}_{-0.61}$ & ${}^{+0.91}_{-0.99}$ & $-1.20$ & ${}^{+0.35}_{-0.50}$ & ${}^{+0.51}_{-0.76}$ & $-1.13$ & ${}^{+0.31}_{-0.54}$ & ${}^{+0.47}_{-0.81}$\\
$\Omega_k$ & $-0.023$ & ${}^{+0.025}_{-0.032}$ & ${}^{+0.088}_{-0.071}$ & $-0.037$& ${}^{+0.024}_{-0.028}$ & ${}^{+0.048}_{-0.065}$ & $-0.037$ & ${}^{+0.024}_{-0.035}$ & ${}^{+0.052}_{-0.070}$\\
$q_0$ & $-0.51$ & ${}^{+0.96}_{-0.70}$ & ${}^{+1.01}_{-1.25}$ & $-0.83$ & ${}^{+0.34}_{-0.47}$ & ${}^{+0.52}_{-0.80}$ & $-0.75$ & ${}^{+0.32}_{-0.54}$ & ${}^{+0.52}_{-0.83}$\\
$\Gamma \equiv \Omega_mh$ & $0.206$ & ${}^{+0.160}_{-0.050}$ & ${}^{+0.182}_{-0.087}$ & $0.190$ & ${}^{+0.030}_{-0.036}$ & ${}^{+0.049}_{-0.065}$ & $0.190$ & ${}^{+0.032}_{-0.034}$ & ${}^{+0.062}_{-0.062}$\\
\hline
\multicolumn{1}{|c|}{$\chi^2/\mathrm{dof}$} & \multicolumn{3}{c|}{$1428/1403$} & \multicolumn{3}{c|}{$1427/1403$} & \multicolumn{3}{c|}{$1427/1403$}\\
\hline
\end{tabular}
\\[0.5cm]
\begin{tabular}{|l|l|l|l|l|l|l|l|l|l|}
\cline{2-10}
\multicolumn{1}{c|}{} & \multicolumn{9}{c|}{7 free parameters: $\Omega_bh^2$, $\Omega_{cdm}h^2$, $\theta$, $\tau$, $n_s$, $A_s$, $\Sigma m_{\nu}$}\\
\cline{2-10}
\multicolumn{1}{c|}{} & \multicolumn{9}{c|}{fixed parameters: $w_{\mathrm{eff}}=-1$, $\Omega_k=0$, $A_T=0$, $\frac{{\mathrm d}\ln n_s}{{\mathrm d}\ln k}=0$}\\
\hline
 & \multicolumn{3}{c|}{WMAP + HST} & \multicolumn{3}{c|}{WMAP + SDSS LRG} & \multicolumn{3}{c|}{WMAP + SDSS LRG}\\
Parameter & \multicolumn{3}{c|}{} & \multicolumn{3}{c|}{sound horizon; adiab. (SH1)} & \multicolumn{3}{c|}{sound horizon (SH2)}\\
\cline{2-10}
 & Median & 68\% & 95\% & Median & 68\% & 95\% & Median & 68\% & 95\%\\
\hline
$H_0 [\mathrm{km/s/Mpc}]$ & $69.2$ & ${}^{+6.4}_{-6.1}$ & ${}^{+13.7}_{-11.2}$ & $69.2$ & ${}^{+2.4}_{-2.5}$ & ${}^{+5.0}_{-5.4}$ & $68.5$ & ${}^{+4.4}_{-3.9}$ & ${}^{+9.0}_{-7.5}$\\
$\Omega_m$ & $0.300$ & ${}^{+0.086}_{-0.066}$ & ${}^{+0.174}_{-0.118}$ & $0.294$ & ${}^{+0.014}_{-0.013}$ & ${}^{+0.028}_{-0.025}$ & $0.302$ & ${}^{+0.044}_{-0.039}$ & ${}^{+0.094}_{-0.072}$\\
$\Sigma m_{\nu} [\mathrm{eV}]$ & ------ & $< 0.80$ & $ < 1.71$ & ------ & $< 0.72$ & $ < 1.63 $ & ------ & $< 0.77 $ & $ < 1.67 $\\
$q_0$ & $-0.54$ & ${}^{+0.13}_{-0.10}$ & ${}^{+0.27}_{-0.18}$ & $-0.552$ & ${}^{+0.024}_{-0.021}$ & ${}^{+0.050}_{-0.041}$ & $-0.539$ & ${}^{+0.069}_{-0.061}$ & ${}^{+0.150}_{-0.112}$\\
$\Gamma \equiv \Omega_mh$ & $0.208$ & ${}^{+0.037}_{-0.034}$ & ${}^{+0.077}_{-0.060}$ & $0.204$ & ${}^{+0.009}_{-0.009}$ & ${}^{+0.018}_{-0.017}$ & $0.207$ & ${}^{+0.021}_{-0.019}$ & ${}^{+0.042}_{-0.035}$\\
\hline
\multicolumn{1}{|c|}{$\chi^2/\mathrm{dof}$} & \multicolumn{3}{c|}{$1429/1404$} & \multicolumn{3}{c|}{$1429/1404$} & \multicolumn{3}{c|}{$1429/1404$}\\
\hline
\end{tabular}
\\[0.5cm]
\begin{tabular}{|l|l|l|l|l|l|l|l|l|l|}
\cline{2-10}
\multicolumn{1}{c|}{} & \multicolumn{9}{c|}{8 free parameters: $\Omega_bh^2$, $\Omega_{cdm}h^2$, $\theta$, $\tau$, $n_s$, $A_s$, $w_{\mathrm{eff}}$, $\Sigma m_{\nu}$}\\
\cline{2-10}
\multicolumn{1}{c|}{} & \multicolumn{9}{c|}{fixed parameters: $\Omega_k=0$, $A_T=0$, $\frac{{\mathrm d}\ln n_s}{{\mathrm d}\ln k}=0$}\\
\hline
 & \multicolumn{3}{c|}{WMAP + HST} & \multicolumn{3}{c|}{WMAP + SDSS LRG} & \multicolumn{3}{c|}{WMAP + SDSS LRG}\\
Parameter & \multicolumn{3}{c|}{} & \multicolumn{3}{c|}{sound horizon; adiab. (SH1)} & \multicolumn{3}{c|}{sound horizon (SH2)}\\
\cline{2-10}
 & Median & 68\% & 95\% & Median & 68\% & 95\% & Median & 68\% & 95\%\\
\hline
$H_0 [\mathrm{km/s/Mpc}]$ & $70.0$ & ${}^{+8.3}_{-7.3}$ & ${}^{+15.6}_{-14.8}$ & $69.2$ & ${}^{+2.6}_{-2.8}$ & ${}^{+5.1}_{-5.3}$ & $68.6$ & ${}^{+4.5}_{-3.6}$ & ${}^{+9.3}_{-7.2}$\\
$\Omega_m$ & $0.298$ & ${}^{+0.093}_{-0.073}$ & ${}^{+0.226}_{-0.122}$ & $0.306$ & ${}^{+0.044}_{-0.051}$ & ${}^{+0.076}_{-0.085}$ & $0.313$ & ${}^{+0.049}_{-0.052}$ & ${}^{+0.100}_{-0.095}$\\
$w_{\mathrm{eff}}$ & $-1.14$ & ${}^{+0.38}_{-0.42}$ & ${}^{+0.66}_{-0.74}$ & $-1.08$ & ${}^{+0.31}_{-0.37}$ & ${}^{+0.45}_{-0.74}$ & $-1.12$ & ${}^{+0.31}_{-0.38}$ & ${}^{+0.48}_{-0.72}$\\
$\Sigma m_{\nu} [\mathrm{eV}]$ & ------ & $< 0.95$ & $ < 1.80$ & ------ & $< 0.95$ & $ < 1.80 $ & ------ & $< 0.95 $ & $ < 1.80 $\\
$q_0$ & $-0.68$ & ${}^{+0.42}_{-0.47}$ & ${}^{+0.80}_{-0.92}$ & $-0.62$ & ${}^{+0.26}_{-0.28}$ & ${}^{+0.39}_{-0.56}$ & $-0.64$ & ${}^{+0.28}_{-0.32}$ & ${}^{+0.46}_{-0.60}$\\
$\Gamma \equiv \Omega_mh$ & $0.209$ & ${}^{+0.040}_{-0.035}$ & ${}^{+0.090}_{-0.063}$ & $0.211$ & ${}^{+0.027}_{-0.031}$ & ${}^{+0.046}_{-0.052}$ & $0.214$ & ${}^{+0.028}_{-0.032}$ & ${}^{+0.052}_{-0.053}$\\
\hline
\multicolumn{1}{|c|}{$\chi^2/\mathrm{dof}$} & \multicolumn{3}{c|}{$1429/1403$} & \multicolumn{3}{c|}{$1429/1403$} & \multicolumn{3}{c|}{$1429/1403$}\\
\hline
\end{tabular}
}
\end{table*}

In this section we investigate the effect of several previously made assumptions and carry out an extended analysis by relaxing some of these. 

We start out with a small comment about tensor modes. There exist several inflationary models which predict a non-negligible tensor mode contribution to the CMB angular power spectrum (for a classification of several models see e.g. \citealt{1997PhRvD..56.3207D}). However, the inclusion of the tensor modes into our analysis would not help to constrain them better than the results obtainable from the CMB data alone \footnote{For the results using the CMB data alone see e.g. \citet{2003ApJS..148..175S}.}. In order to make a clean separation of the tensor and scalar contributions to the CMB angular fluctuations one would greatly benefit from the independent knowledge of the scalar fluctuation level as obtainable from the large-scale structure studies. Nevertheless, as in our analysis the biasing parameter is treated as a completely free quantity that is marginalized out, we do not have any sensitivity to the absolute level of the scalar fluctuation component. It is worth pointing out that in principle a good handle on bias parameter can be obtained by the study of the higher order clustering measures e.g. bispectrum \citep{1997MNRAS.290..651M,1998MNRAS.300..747V}.    

Since several neutrino oscillation experiments unambiguously indicate that neutrinos have a non-zero mass (for a recent review see e.g. \citealt{2006PhR...429..307L}) it would be interesting to investigate models with massive neutrino component. For simplicity we concentrate on models with three generations of neutrinos with degenerate masses. Also, as the generic prediction of almost all the inflationary models is the nonmeasurably small spatial curvature i.e. $\Omega_k \simeq 0$ it is of great interest to test whether this is compatible with the observational data. For these reasons we have extended our initial analysis to allow for the massive neutrino component and also have investigated the effect of relaxing the assumption about spatial flatness. The results of this study for some of the model classes are briefly presented in Table \ref{tab2}. Here we have not carried out an analysis using the full measured LRG power spectrum, instead only the measurement of the acoustic scale was added as an additional information to the CMB data. This can be justified for two reasons: (i) To model the observed power spectrum correctly one needs to incorporate the corrections due to nonlinearities and redshift-space distortions, which in our case was done simply by introducing one additional parameter $Q$ (see Eq. (\ref{eq1})). However, it is clear that the effect of this parameter can resemble very closely the power-damping effect of massive neutrinos i.e. there will be very strong degeneracies that result in poor constraints on neutrino mass. Instead of the dynamic damping effect of massive neutrinos we can probe their effect on the kinematics of the background expansion by exploiting the measurement of the low redshift acoustic scale. (ii) The non-zero spatial curvature has twofold effect on the matter power spectrum. First, it influences the growth rate of the fluctuations and thus changes the amplitude of the spectrum. Since in our case the amplitude is treated as a free parameter we are unable to use this effect. Second, the spatial scales get transformed, resulting in the horizontal stretch/compression of the power spectrum. It is clear that the smooth ``continuum'' of the power spectrum with an unknown amplitude does not contain much information about the possible horizontal stretch/compression \footnote{That is especially true in the case we do not have information about the scale of the turnover in the spectrum (see Fig. \ref{fig1} lower panel).} as this transformation can be easily mimicked by the corresponding change of the normalization. The degeneracy between these two transformations can be broken if the spectrum contains some sharper features e.g. acoustic oscillations.

\begin{figure}
\centering
\includegraphics[width=\plotwd]{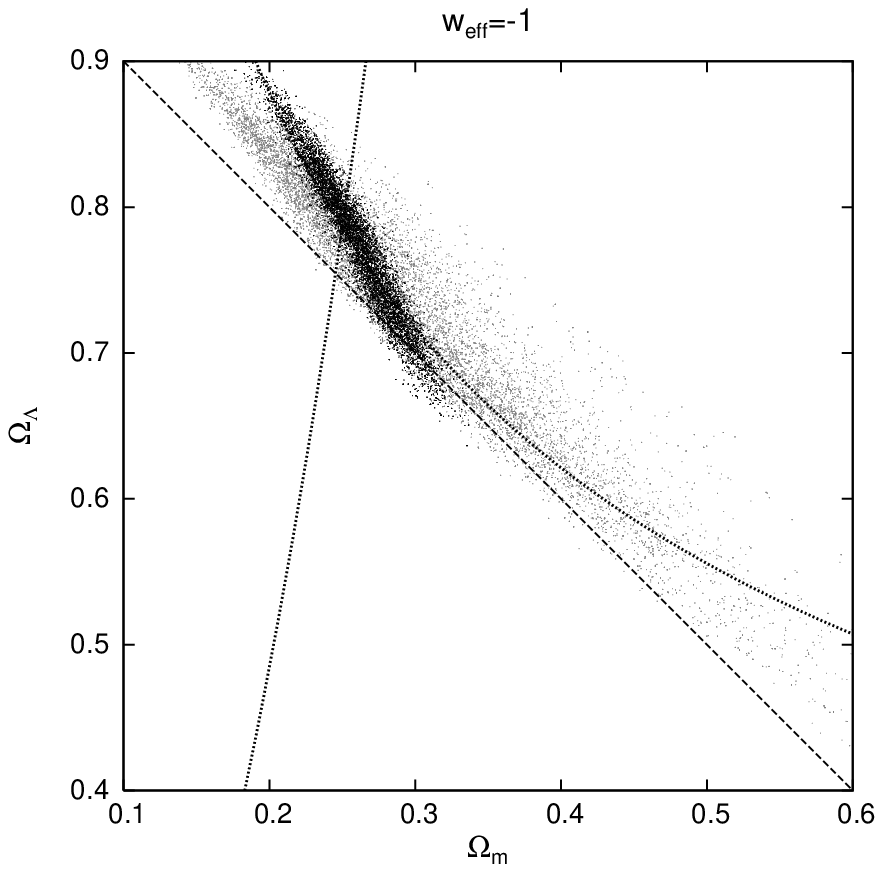}
\caption{The distribution of models in $\Omega_m-\Omega_{\Lambda}$ plane. Gray points correspond to the {\sc Wmap}+HST case, black ones also include the low redshift sound horizon measurement. The dashed line represents spatially flat models. Dotted lines provide approximate degeneracy directions as given by Eqs. (\ref{deg1}) and (\ref{deg2}). Dark energy is assumed to be given by the cosmological constant.}
\label{fig15_5}
\end{figure}

According to the results presented in Fig. \ref{fig17} one probably would not expect any significant improvement on the measurement of the neutrino mass over the one obtained using the {\sc Wmap} data alone. This is indeed the case as can be seen from Table \ref{tab2}. Our results on $\Sigma m_{\nu}$ are in good agreement with the constraints obtained in \citet{2005PhRvD..71d3001I,2006PhRvD..74d7302F}. Contrary to the results presented in \citet{2005ApJ...633..560E} we do not find any improvement in the measurement of $\Omega_k$ once the measurement of the low redshift acoustic scale is incorporated into the analysis. Although by measuring more accurately the value of the Hubble constant one would expect to better break the geometric degeneracy \citep{1997MNRAS.291L..33B} and thus should in principle get better constraint on $\Omega_k$, this is currently not the case. As seen from Fig. \ref{fig15_5}, by including acoustic scale information, model points lie inside the significantly reduced ellipse that is tilted with respect to the bigger one that uses the {\sc Wmap}+HST data alone, in such a way, that the projection perpendicular to the flatness line still has practically the same width i.e. the constraints on $\Omega_k$ are also practically identical. We can see how the original distribution of the models turns towards the and also gets significantly compressed perpendicular to the almost vertical degeneracy line corresponding to the measurement of the low redshift sound horizon (see Eq. (\ref{deg2})). In Table \ref{tab2} along with $\Sigma m_{\nu}$ and $\Omega_k$ we have also given a subset of parameters that benefit mostly from the inclusion of the measurement of the acoustic scale. As expected, the parameter constraints are generally getting weaker as we allow for more freedom in the models.   

\section{Discussion and Conclusions}
In this paper we have performed a MCMC cosmological parameter study using the results from the recent SDSS LRG power spectrum analysis by \citet{astro-ph/0512201} along with the CMB temperature-temperature and temperature-polarization angular power spectra as determined by the {\sc Wmap} team \citep{2003ApJS..148..135H,2003ApJS..148..161K}. We have carried out the analysis in two parts: (i) using the {\sc Wmap} data + the measurement of the low redshift sound horizon as found from the SDSS LRG redshift-space power spectrum, (ii) using the {\sc Wmap} data + full SDSS LRG power spectrum as shown in Fig. \ref{fig1}. As the formation of the acoustic features in the large-scale matter distribution is theoretically very well understood the separate treatment for the oscillatory part of the LRG power spectrum is well justified. Moreover, in comparison to the full power spectrum, which along with the dependence on several cosmological parameters requires additional modeling of the redshift-space distortions, nonlinear evolution, and biasing \footnote{Assuming we do not want to exclude the quasilinear scales from our analysis.}, the acoustic scale depends on only a few cosmological parameters. The CMB measurements calibrate the physical scale of the sound horizon to a very good accuracy. By comparing it with the scale inferred from the low redshift LRG power spectrum measurements, we are able to get a very tight constraint on the Hubble parameter: $h=0.708^{+0.021}_{-0.020}$ if assuming adiabatic initial conditions, or $h=0.705^{+0.038}_{-0.037}$ if additional shift in oscillation phase is allowed. Having a tight constraint on $h$ allows us to break several parameter degeneracies, and thus helps us to determine various parameters like $\Omega_m$, $\Omega_{cdm}$, $\Omega_{b}$ with a good precision. Also, in comparison to the {\sc Wmap} + HST data, a significantly tighter constraint on $w_{\mathrm{eff}}$ is obtained. The full results for all the parameters are summarized in Table \ref{tab1}. The obtained values are in general in a good agreement with several other parameter studies e.g. \citet{2002MNRAS.337.1068P,2003ApJS..148..175S,2004PhRvD..69j3501T}. Relatively tight bounds on ($H_0$, $\Omega_m$, $w_{\mathrm{eff}}$) or equivalently on ($H_0$, $q_0$, $j_0$) help us to determine the low redshift expansion law with significantly higher precision than available from the {\sc Wmap} + HST data alone. If the initial fluctuations are constrained to be adiabatic, the measurement of the acoustic scale rules out a decelerating Universe, i.e. $q_0>0$, at $5.5\sigma$ confidence level. 

In contrast to the acoustic scale measurement, that gave a precise value for the Hubble parameter, the full spectrum provides us with a good estimate for the shape parameter $\Gamma\equiv\Omega_mh=0.207^{+0.011}_{-0.012}$, which is in a very good agreement with the one found in \citet{2004PhRvD..69j3501T}. Since in the $\Omega_m-h$  plane the $\Gamma\equiv\Omega_mh=\mathrm{const}$ line (see Fig. \ref{fig13}) is only relatively weakly tilted with respect to the relevant CMB degeneracy direction $\Omega_mh^2=\mathrm{const}$, the obtained limits on $\Omega_m$ and $h$ are not as strong as the ones obtained from the measurement of the acoustic scale. In contrast, the degeneracy lines corresponding to the low redshift acoustic scale measurement are in many cases almost orthogonal to the {\sc Wmap} + HST ``ellipses'', which explains the stronger constraints for several parameters. 

Throughout most of this work we have focused on spatially flat models and also have assumed negligible contribution from massive neutrinos. As expected, by relaxing these assumptions the bounds on several cosmological parameters loosen significantly. The results of this extended analysis for some of the parameters are represented in a compact form in Table \ref{tab2}.

We have also stressed the need to apply cosmological transformations to the theoretical model spectra before being compared with the relevant observational spectrum, which is valid only in the reference frame of the fiducial cosmological model that was used to analyze the data. So far almost all the parameter studies have completely ignored this point, which is probably fine for the shallow redshift surveys. On the other hand, in case of more deeper surveys like the SDSS LRG, reaching $z\sim 0.5$, these transformations have to be certainly applied. In general the line intervals along and perpendicular to the line of sight transform differently. Also the transformations depend on redshift. We have shown that for the samples like SDSS LRGs, with a typical redshift of $z\sim 0.35$, the single ``isotropized'' transformation taken at the median redshift of the survey provides a very accurate approximation to the more complete treatment.

For the parameter estimation we have used the SDSS LRG power spectrum down to the quasilinear scales, which calls for the extra treatment of nonlinear effects, small scale redshift-space distortions and biasing. These additional complications can be relatively well dealt with the aid of the Halo Model (see Appendix \ref{appa}). We have shown in \citet{astro-ph/0512201} that a simple analytical model with additional four free parameters is able to approximate the observed spectrum to a very good precision. Also, the Halo Model has been shown to provide a good match to the results of the semianalytical galaxy formation studies (see e.g. \citealt{2002PhR...372....1C}). In this paper we have shown that to a rather tolerable accuracy the above four extra parameter Halo Model spectra (for reasonable values of the parameters) can be represented as a simple transformation of the linear power spectrum with only one extra parameter (see Fig. \ref{fig2}). The similar type of transformation was also used in \citet{2005MNRAS.362..505C}.

In order to investigate the possible biases introduced by the specific method used to extract the sound horizon from the power spectrum measurements, we have performed a Monte Carlo study, the results of what are shown in Fig. \ref{fig7}.

\acknowledgements{I thank Enn Saar and Jose Alberto Rubi\~no-Mart\'in for valuable discussions and Rashid Sunyaev for the comments on the manuscript. Also I am grateful to the referee for helpful comments and suggestions. I acknowledge the Max Planck Institute for Astrophysics for a graduate fellowship and the support provided through Estonian Ministry of Education and Recearch project TO 0062465S03.}

\bibliographystyle{aa}

\appendix

\section{Power spectrum from the halo model}\label{appa}
The halo model description of the spatial clustering of galaxies is a development of the original idea by \citet{1952ApJ...116..144N}, where one describes the correlations of the total point set as arising from two separate terms: (i) the 1-halo term, that describes the correlations of galaxies populating the same halo, (ii) the 2-halo term, which takes into account correlations of the galaxies occupying different halos. For a thorough review see \citet{2002PhR...372....1C}. Here we briefly give the results relevant to the current paper (see \citealt{2001MNRAS.325.1359S,2004MNRAS.348..250C}).

The power spectrum of galaxies in redshift space can be given as:
\begin{equation}
P(k) = P^{1h}(k) + P^{2h}(k)\,,
\end{equation}
where the 1-halo term:
\begin{eqnarray}
P^{1h}(k) & = & \int \mathrm{d}M \,n(M)\frac{\langle N(N-1)|M \rangle}{\bar{n}^2}\mathcal{R}_p(k\sigma)|u_g(k|M)|^p\,, \\
p & = &\left\{ 
\begin{array}{ll}
1 & \mathrm{\quad if \quad} \langle N(N-1)|M \rangle < 1\\
2 & \mathrm{\quad if \quad} \langle N(N-1)|M \rangle > 1
\end{array}
\right.
\end{eqnarray}
and the 2-halo term:
\begin{equation}
P^{2h}(k) = \left ( \mathcal{F}_g^2 + \frac{2}{3}\mathcal{F}_v\mathcal{F}_g + \frac{1}{5}\mathcal{F}_v^2 \right )P_\mathrm{lin}(k)\,. 
\end{equation}
Here:
\begin{eqnarray}
& & \mathcal{R}_p \left ( \alpha  = k \sigma \sqrt {\frac{p}{2}} \right ) =  \frac{\sqrt{\pi}}{2}\frac{\mathrm{erf}(\alpha)}{\alpha}\,, \label{eq_b5} \\ 
& & \mathcal{F}_g  =  \int \mathrm{d}M \,n(M) b(M) \frac{\langle N|M \rangle}{\bar{n}}
\mathcal{R}_1(k\sigma) u_g(k|M)\,, \\
& & \mathcal{F}_v =  f \cdot \int \mathrm{d}M\,n(M) b(M) \mathcal{R}_1(k\sigma) u(k|M)\,. \label{eq_b1}
\end{eqnarray}
In the above expressions $n(M)$ is the mass function and $b(M)$ halo bias parameter. We calculate them using the prescription by \citet{1999MNRAS.308..119S} and \citet{2001MNRAS.323....1S}. $\bar{n}$ represents the mean number density of galaxies:
\begin{equation}
\bar{n} = \int \mathrm{d}M\, n(M) \langle N|M \rangle\,. 
\end{equation}
We take the mean of the halo occupation distribution in the following form:
\begin{equation}\label{eq_b2}
\langle N|M \rangle = \left ( \frac{M}{M_0} \right )^\alpha\,,
\end{equation}
where $M_0$ and $\alpha$ are free parameters. The second moment is chosen as (see \citealt{2004MNRAS.348..250C}):
\begin{eqnarray}
& & \langle N(N-1)|M \rangle  =  \beta^2(M) \langle N|M \rangle ^2\,, \\
& & \beta(M)  =  \left \{ 
\begin{array}{ll}
\frac{1}{2} \log \left ( \frac{M}{10^{11} \ h^{-1}M_{\odot}} \right ) & \mathrm{\quad if \quad} M<10^{13} \  h^{-1}M_{\odot} \\
1 & \mathrm{\quad otherwise.}\label{eq_b4}
\end{array}
\right.
\end{eqnarray}
$f$ in Eq. (\ref{eq_b1}) denotes the logarithmic derivative of the linear growth factor: $f \equiv \frac{\mathrm{d} \ln D_1}{\mathrm{d} \ln a}$. $u(k|M)$ and $u_g(k|M)$ are the normalized Fourier transforms of the dark matter and galaxy density distributions within a halo of mass $M$. In our calculations we take both of these distributions given by the NFW profile \citep{1997ApJ...490..493N} and the concentration parameter $c(M)$ is taken from \citet{2001MNRAS.321..559B}. The one dimensional velocity dispersion of the galaxies inside a halo with mass $M$ is taken to follow the scaling of the isothermal sphere model:
\begin{equation}\label{eq_b3}
\sigma = \gamma \sqrt {\frac{GM}{2R_\mathrm{vir}}}\,,
\end{equation}
where $R_\mathrm{vir}$ is the virial radius of the halo and $\gamma$ is a free parameter.  

After specifying the background cosmology the above described model has four free parameters: $M_0$, $\alpha$ (Eq. (\ref{eq_b2})), $\sigma$ (Eq. (\ref{eq_b3})) and $M_\mathrm{low}$. The last parameter $M_\mathrm{low}$ represents the lower boundary of the mass integration i.e. halos with masses below $M_\mathrm{low}$ are assumed to be ``dark''.

\section{Fitting formulae for the acoustic scales}\label{appb}
The comoving distance traveled by the sound wave since the Big Bang up to redshift $z$ can be expressed as:
\begin{equation}\label{eq1_23}
s(z)=\int\limits_{\infty}^{z}c_s(z')(1+z')\frac{{\mathrm d}t}{{\mathrm d}z'}{\mathrm d}z'\,,
\end{equation}
where the sound speed:
\begin{eqnarray}
c_s(z)&=&\frac{c}{\sqrt{3\left[1+\mathcal{R}(z)\right]}}\,,\\
\mathcal{R}(z)&\equiv&\frac{3\rho_b}{4\rho_{\gamma}}\simeq 3.04\times 10^4\cdot\frac{\Omega_b h^2}{z}\,.
\end{eqnarray}
Using the Friedmann equation, Eq. (\ref{eq1_23}) can be integrated to yield (see e.g. \citealt{astro-ph/9508126})\footnote{The result is valid for high enough redshifts as relevant for the propagation of the sound waves.}:
\begin{eqnarray}
s(z) & = & \frac{3.46\times 10^3\, \mathrm{Mpc}}{\sqrt{\Omega_m h^2\cdot z_{\mathrm{eq}}\mathcal{R}(z_{\mathrm{eq}})}}\cdot \nonumber \\
& & \ln\left[\frac{\sqrt{1+\mathcal{R}(z)}+\sqrt{\mathcal{R}(z)+\mathcal{R}(z_{\mathrm{eq}})}}{1+\sqrt{\mathcal{R}(z_{\mathrm{eq}})}}\right]\,, \label{eq1_26}
\end{eqnarray}
where the redshift for the matter-radiation equality can be given as:
\begin{equation}\label{eq1_27}
z_{\mathrm{eq}}\simeq 2.41\times 10^4\cdot\Omega_m h^2\,.
\end{equation}
The acoustic scale relevant for the CMB studies is $s_*=s(z_*)$, where $z_*$ denotes the recombination redshift. For the ``concordance'' cosmological model the acoustic scale imprinted in the matter power spectrum $s_d=s(z_d)$ is slightly larger than $s_*$. Here $z_d$ denotes the redshift at which the baryons are released from the Compton drag of the photon field. To find accurate values for $z_*$ and $z_d$ one has to carry out a full calculation for the recombination history of the Universe. The results of these calculations can be conveniently expressed as the following fitting formulas (accurate at $\sim 1\%$ level) \citep{1996ApJ...471..542H,1998ApJ...496..605E}:
\begin{equation}
z_*=1048\left[1+0.00124(\Omega_bh^2)^{-0.738}\right]\left[1+g_1(\Omega_mh^2)^{g_2}\right]\,,
\end{equation}
where
\begin{eqnarray}
g_1&=&0.0783(\Omega_bh^2)^{-0.238}\left[1+39.5(\Omega_bh^2)^{0.763}\right]^{-1}\,,\\
g_2&=&0.560\left[1+21.1(\Omega_bh^2)^{1.81}\right]^{-1}\,,
\end{eqnarray} 
and
\begin{eqnarray}
z_d &=& 1291(\Omega_mh^2)^{0.251}\left[1+0.659(\Omega_mh^2)^{0.828}\right]^{-1}\cdot\nonumber \\
& & \left[1+b_1(\Omega_bh^2)^{b_2}\right]\,,\label{eq1_30}
\end{eqnarray}
where
\begin{eqnarray}
b_1&=&0.313(\Omega_mh^2)^{-0.419}\left[1+0.607(\Omega_mh^2)^{0.674}\right]\,,\label{eq1_31}\\
b_2&=&0.238(\Omega_mh^2)^{0.223}\,.\label{eq1_32}
\end{eqnarray} 
For $\Omega_bh^2\lesssim 0.03$ the drag epoch follows the last scattering of the photons.

\begin{figure}
\centering
\includegraphics[width=\plotwd]{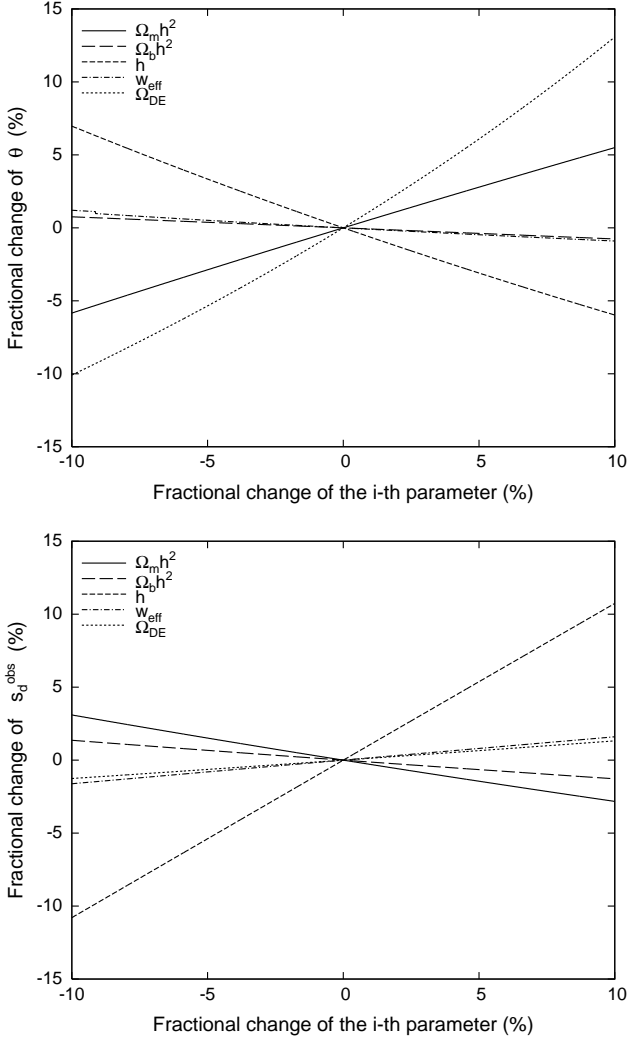}
\caption{The dependence of the angular scale corresponding to the sound horizon at decoupling (upper panel), and the sound horizon as measured from the low redshift matter power spectrum (lower panel), on various cosmological parameters. In the lower panel an effective observational redshift $z_{\mathrm{eff}}=0.35$ has been assumed. The variation of the parameters has been performed around the ``concordance'' cosmological model. For this model the central values for $\theta$ and $s_d^{\mathrm{obs}}$ are $\sim 0.6^{\circ}$ and $\sim 107\,h^{-1}\,\mathrm{Mpc}$, respectively.}
\label{fig1_7}
\end{figure}

From the CMB measurements one can determine the angular scale $\theta$ that corresponds to the sound horizon at decoupling with a good accuracy. This angle can be expressed as:
\begin{equation}
\theta = \frac{s_*h}{d_{\perp}(z_*)}\,,
\end{equation}
where $s_*=s(z_*)$, as given in Eq. (\ref{eq1_26}), is measured in $\mathrm{Mpc}$, and we have added an extra factor of $h$ to the numerator to convert to the usual units of $h^{-1}\,\mathrm{Mpc}$. Here $d_{\perp}(z_*)$ is the comoving angular diameter distance to the last scattering surface, which is strongly dependent on the curvature radius $R_0$. As $s_*$ is only weakly dependent on $\Omega_mh^2$ and $\Omega_bh^2$, it turns out that measurement of $\theta$ is very sensitive to $\Omega_k$. The dependence of $\theta$ on various cosmological parameters (around the ``concordance'' model point) is given in the upper panel of Fig. \ref{fig1_7}. Using the following set of model parameters: ($\Omega_mh^2$, $\Omega_bh^2$, $h$, $\Omega_{DE}$, $w_{\mathrm{eff}}$), the measurement of $\theta$ constrains directly the linear combination 
\[
0.40\left(\frac{\Omega_mh^2}{0.14}\right) + 0.80\left(\frac{\Omega_{DE}}{0.73}\right) - 0.45\left(\frac{h}{0.71}\right)\, ,\nonumber 
\]
or in case of logarithmic variables the combination
\begin{equation}\label{deg1}
(\Omega_mh^2)^{0.40}(\Omega_{DE})^{0.80}(h)^{-0.45}\,.
\end{equation}

To measure the sound horizon $s_d$ in the large-scale matter distribution one has to assume some background cosmological model in order to convert the observed redshifts to comoving distances. If the assumed fiducial model differs from the true cosmology, the measured scale will also be distorted. As shown in Chapter 4, for relatively low redshift measurements this distortion can be approximated as a single transformation for the ``isotropized'' comoving interval. (In general the comoving intervals along and perpendicular to the line of sight transform differently, and these transformations also depend on redshift.) The observed sound horizon $s_d^{\mathrm{obs}}$ can be approximated as:
\begin{equation}
s_d^{\mathrm{obs}}=c_{\mathrm{isotr}}\cdot hs_d=\sqrt[3]{c_{\parallel}c_{\perp}^2}\cdot hs_d\,, 
\end{equation}
where the extra factor of $h$ is again included to convert to $h^{-1}\,\mathrm{Mpc}$, and the functions $c_{\parallel}(z)$, $c_{\perp}(z)$, which should be evaluated at the effective redshift $z_{\mathrm{eff}}$ of the observations (e.g. median redshift), are defined as:
\begin{equation}
c_{\parallel}(z) = \frac{E^{\mathrm{fid}}(z)}{E(z)},\quad c_{\perp}(z) = \frac{d_{\perp}(z)}{d_{\perp}^{\mathrm{fid}}(z)}\,,
\end{equation}
where
\begin{eqnarray}
E(z) &\equiv& \Bigg[ \Omega_r(1+z)^4 + \Omega_m(1+z)^3 + \Omega_k(1+z)^2 + \Bigg. \nonumber \\
& & \Bigg. + \Omega_{DE}\exp \int \limits_{0}^{\ln(1+z)}3\left[1+w_{DE}(z')\right]{\mathrm d}z'\Bigg]^{\frac{1}{2}}\,, \label{eq15}\\
d_{\perp}(z) &=& R_0S_k\left(\frac{d_{\parallel}(z)}{R_0}\right)\,,\, S_k(x)=\left\{
\begin{array}{lll}
\sin x &  \mathrm{if} \ \Omega_k < 0 \\
x &  \mathrm{if} \ \Omega_k = 0 \\
\sinh x & \mathrm{if} \ \Omega_k > 0 \ ,
\end{array} \right.,\\
\quad R_0 & = & \frac{d_H}{\sqrt{|\Omega_k|}}\,, \quad d_H = \frac{c}{H_0}\,,\quad d_{\parallel}(z) = d_H\int\limits_{0}^{z}\frac{{\mathrm d}z'}{E(z')}\,.
\end{eqnarray}
The superscript $^{\mathrm{fid}}$ refers to the fiducial model. The dependence of $s_d^{\mathrm{obs}}$ on various cosmological parameters is shown in the lower panel of Fig. \ref{fig1_7}. The fiducial model here was taken to be the best-fit {\sc Wmap} cosmology and the ``true models'' were assumed to populate its intermediate neighborhood. We also assume $z_{\mathrm{eff}}=0.35$ as in the case of the SDSS LRG sample analyzed in this thesis. Then the linear combination of parameters constrained by the measurement of $s_d^{\mathrm{obs}}$ turns out to be
\[
-0.26\left(\frac{\Omega_mh^2}{0.14}\right)-0.11\left(\frac{\Omega_bh^2}{0.022}\right)+0.94\left(\frac{h}{0.71}\right)+0.12\left(\frac{\Omega_{DE}}{0.73}\right)+0.15\left(\frac{w_{\mathrm{eff}}}{-1}\right)\,,
\]
or in case of logarithmic variables
\begin{equation}\label{deg2}
(\Omega_mh^2)^{-0.26}(\Omega_bh^2)^{-0.11}(h)^{0.94}(\Omega_{DE})^{0.12}(w_{\mathrm{eff}})^{0.15}\,.
\end{equation}
Thus, as probably expected, the strongest dependence is on the Hubble parameter $h$ i.e. the precise measurement of the acoustic scale at low redshifts should give us a good estimate for $h$.   
\subsection{The effect of massive neutrinos}
In addition to the dynamical effect that massive neutrinos have on the evolution of density perturbations they also lead to the modification of the expansion law of the Universe. Assuming three families of massive neutrinos with degenerate masses $m_{\nu}$ one has to replace the term $\Omega_r(1+z)^4$ in Eq. (\ref{eq15}) with:
\begin{eqnarray}
\Omega_{\gamma}(1 &+& z)^4\left [1 + 0.1199 \cdot I(z) \right ],\, \mathrm{where:}\\
I(z) &=& \int \limits_{0}^{\infty}\sqrt{x^2+f^2(z)}\cdot\frac{x^2}{e^x + 1} {\mathrm d}x,\,\label{eq16}\\
f(z) &=& \frac{5967\cdot m_{\nu} [\mathrm{eV}]}{1+z} \,.
\end{eqnarray}
Here the density parameters corresponding to the photons $\Omega_{\gamma}$ and total radiation $\Omega_r$ are related by $\Omega_r=1.68\cdot\Omega_{\gamma}$.

\begin{figure}
\centering
\includegraphics[width=\plotwd]{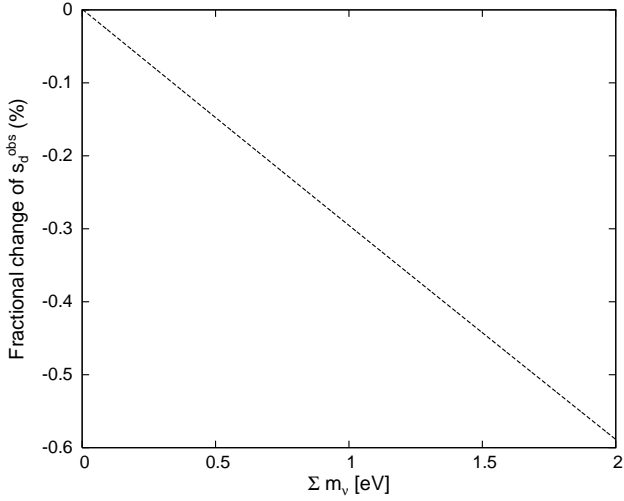}
\caption{The relative change of the sound horizon, as measured from the low redshift matter power spectrum, as a function of the sum of neutrino masses. Here an effective observational redshift $z_{\mathrm{eff}}=0.35$ has been assumed. The other parameters are kept fixed to the best-fit {\sc Wmap} $\Lambda$CDM model values. }
\label{fig17}
\end{figure}

Since the lower limit for the sum of neutrino masses coming from the oscillation experiments $\Sigma m_{\nu}\gtrsim 0.056\, \mathrm{eV}$ (e.g. \citealt{2006PhR...429..307L}) is large enough, Eq. (\ref{eq16}) at low redshifts (as relevant for the acoustic scale measurements using the galaxy clustering data) can be very well approximated as:
\begin{equation}
I(z)=\frac{1.076\times 10^4 \cdot m_{\nu}\,[\mathrm{eV}]}{1+z}\,,
\end{equation}
i.e. neutrinos act as nonrelativistic matter. In case of the experimentally lowest allowable $m_{\nu}$ this approximation is good to $0.1 \%$ for redshifts $z \lesssim 3.3$.

The relative change of the size of the measurable sound horizon at $z=0.35$ as a function of the sum of neutrino masses $\Sigma m_{\nu}$, for the mass range compatible with the constraints arising from the {\sc Wmap} data \citep{2005PhRvD..71d3001I,2006PhRvD..74d7302F}, is shown in Fig. \ref{fig17}. The other parameters here are kept fixed to the best-fit {\sc Wmap} $\Lambda$CDM model values \citep{2003ApJS..148..175S}. As we can see the measurable sound horizon is only a weak function of $\Sigma m_{\nu}$ and thus one would not expect any strong constraints on neutrino mass from the measurement of the low redshift acoustic scale. We remind that the relative accuracy of the sound horizon measurement as found in \citet{astro-ph/0512201} is in the range $2-7 \%$.    

\section{Relation between ($\Omega_m$, $w_{\mathrm{eff}}$) and ($q_0$, $j_0$)}\label{appc}
In case of the constant dark energy equation of state parameter $w_{\mathrm{eff}}$, deceleration parameter $q_0$ and jerk $j_0$ at redshift $z=0$ can be expressed as:
\begin{eqnarray}
q_0&=&\frac{1}{2}\Omega_m+\frac{1+3w_{\mathrm{eff}}}{2}\Omega_{DE}\,,\\
j_0&=&\Omega_m + \left[1+\frac{9}{2}w_{\mathrm{eff}}(w_{\mathrm{eff}}+1)\right]\Omega_{DE}\,.
\end{eqnarray}
For the spatially flat models i.e. $\Omega_m + \Omega_{DE}=1$ these relations can be uniquely inverted to provide:
\begin{eqnarray}
\Omega_m &=& \frac{2\left[j_0-q_0(1+2q_0)\right]}{1+2(j_0-3q_0)}\,,\\
w_{\mathrm{eff}} &=& \frac{2(3q_0-j_0)-1}{3(1-2q_0)}\,.
\end{eqnarray}

\end{document}